\documentclass{emulateapj}

\def\lsim{\mathrel{\rlap{\lower4pt\hbox{\hskip1pt$\sim$}}\raise1pt\hbox{$<$}}}
\def\gsim{\mathrel{\rlap{\lower4pt\hbox{\hskip1pt$\sim$}}\raise1pt\hbox{$>$}}}

\usepackage{graphicx}

\slugcomment{submitted to the Astrophysical Journal}
\shorttitle{3C 264}
\shortauthors{Perlman et al.}
 
\begin{document}

\title{A Multiwavelength Spectral and Polarimetric Study of the Jet of 3C 264}

\author{E. S. Perlman\altaffilmark{1,2},
C. A. Padgett\altaffilmark{2,3}, 
M. Georganopoulos\altaffilmark{2,3}, 
D. M. Worrall\altaffilmark{4}, 
J. H. Kastner\altaffilmark{5}, 
G. Franz\altaffilmark{5}, 
M. Birkinshaw\altaffilmark{4}, 
F. Dulwich\altaffilmark{4}, 
C. P. O'Dea\altaffilmark{5}, 
S. A. Baum\altaffilmark{6}, 
W. B. Sparks\altaffilmark{7}, 
J. A. Biretta\altaffilmark{7}, 
L. Lara\altaffilmark{8}, 
S. Jester\altaffilmark{9}, 
A. Martel\altaffilmark{7}}

\altaffiltext{1}{Physics and Space Sciences Department, Florida Institute of
Technology, 150 West University Boulevard, Melbourne, FL 32901, USA}

\altaffiltext{2}{Department of Physics, University of Maryland-Baltimore
County, 1000 Hilltop Circle, Baltimore, MD 21250, USA.}

\altaffiltext{3}{NASA's Goddard Space Flight Center, Mail Code 660, Greenbelt, 
MD 20771, USA}

\altaffiltext{4}{Department of Physics, University of Bristol, Tyndall Avenue, 
Bristol BS8 1TL, UK}

\altaffiltext{5}{Physics Department, Rochester Institute of Technology, 84 Lomb
Memorial Drive, Rochester, NY  14623-5603, USA}

\altaffiltext{6}{Center for Imaging Science, Rochester Institute of Technology,
54 Lomb Memorial Drive, Rochester, NY 14623-5603, USA}

\altaffiltext{7}{Space Telescope Science Institute, 3700 San Martin Drive,
Baltimore, MD 21218, USA}

\altaffiltext{8}{Dpto. Fis\'ica Te\'orica y del Cosmos, Universidad de Granada,
Avenida Fuentenueva s/n, 18071 Granada, Spain}

\altaffiltext{9}{Max Planck Institut fur Astronomie, K\"onigstuhl 17, D-69117
Heidelberg, Germany}

\email{eperlman@fit.edu}

\begin{abstract}
We present a comprehensive multiband spectral and polarimetric study of the jet
of 3C 264 (NGC 3862). Included in this study are three {\it HST} optical
and ultraviolet polarimetry data sets, along with new and archival {\it VLA}
radio imaging and polarimetry, a re-analysis of numerous {\it HST} broadband
data sets from the near infrared to the far ultraviolet, and a {\it
Chandra} ACIS-S observation.  We investigate similarities and differences
between optical and radio polarimetry, in both degree of polarization and
projected magnetic field direction. 
We also examine the broadband spectral energy distribution of both the 
nucleus and jet of 3C 264, from the radio through the X-rays.  
From this we place constraints on the physics of the 3C 264 system, the jet and
its dynamics.  We find significant curvature of the spectrum from the near-IR 
to ultraviolet, and synchrotron breaks steeper than 0.5, a situation also
encountered in the jet of M87.  This likely indicates velocity and/or magnetic 
field gradients and more efficient particle acceleration localized in the
faster/higher magnetic field parts of the flow.  
The magnetic field structure of the 3C 264 jet is
remarkably smooth; however, we do find complex magnetic field structure that
is correlated with changes in the optical spectrum.  We find that the 
X-ray emission is due to the synchrotron process; we model the jet spectrum 
and discuss mechanisms for accelerating particles to the needed energies,
together with implications for
the orientation of the jet under a possible spine-sheath model.

\end{abstract}
\maketitle
\vfill\eject

\section{Introduction} 

The characterization of jet polarization properties provides a powerful
diagnostic of jet physics, particularly with respect to magnetic field
configuration and particle acceleration.  Extragalactic jets generally emit a
continuum of radiation from radio through optical, and often into the X-ray
regime. Through matched resolution comparisons of flux density measurements at
various frequencies, we can glean morphological information  about particle
acceleration regions and jet energetic structure.

The closest kiloparsec-scale radio-optical jet is that of M87 at a distance of
$16{\rm ~Mpc}$ (Tonry 1991).  By using matched resolution polarimetry at
different wavelengths and combining this with multi-wavelength imaging and X-ray
imaging and spectroscopy, much information about jet energetics and magnetic
fields, three dimensional structure and particle acceleration can be constrained
(Perlman \& Wilson 2005 and references therein). There are relatively few jets
for which there exists HST optical polarimetry compared to the number that have
been detected.  As of this writing, a total of ten optical jets have HST
polarimetry observations (M87 - e.g. Perlman et al. 1999; 3C 273 - Thomson et
al. 1993; 3C 293 - Floyd et al.  2006; 3C 15 - Perlman et al. 2006 and 
Dulwich et al. 2007; 3C 346 - Perlman et al. 2006 and Dulwich et al. 2009; 
3C 66B, 3C 78, 3C 264 and
3C 371 - Perlman et al. 2006;  PKS 1136--135 - Cara et al., in prep.).   This
is in comparison to the $\sim 34$ detected optical extragalactic jets
{\footnote{see http://home.fnal.gov/\textasciitilde jester/optjets/ maintained by S. Jester}}.
Given this dearth of polarimetry observations, it is not surprising that there
are  few constraints on the configuration of the magnetic fields in optically
emitting regions of extragalactic jets. Here we present a comprehensive study of
the jet of  3C 264 that includes radio and optical polarimetry, as well as X-ray
observations of the jet.

At a redshift of 0.0217 (Baum et al. 1990), and hence a distance of 94 Mpc,  3C
264 is among the closest known bright radio galaxies with an optical jet. Also
detected at X-ray energies, its relative proximity makes the 3C 264 jet a prime
candidate for deep optical, radio and X-ray studies, as we present here.  It is
hosted by NGC~3862, a large elliptical galaxy offset to the south-east from the
center of the cluster Abell 1367, and is classified as a Fanaroff-Riley type I
radio source (Fanaroff \& Riley, 1974). On large scales it exhibits a twin-tail
radio structure extending to the north-east (Bridle \& Vallee 1981). On arcsec
scales it consists of a compact core, with a one-sided, nearly knot-free jet
extending also to the north-east and a weak counter jet (Lara et al. 1997). The
optical jet counterpart extends for only roughly $2''$ beyond the galaxy core,
the latter arcsecond of which is considerably fainter than the inner jet
(Perlman et al. 2006). This short length combined with the ``smoothness'' of the
jet makes studying this object difficult.  In particular, in the optical and
near-IR  useful observations are really only possible with HST, as most
adaptive-optics systems require a nearby, bright point source.

3C 264 was part of the original Third Cambridge catalog with $S_{\rm 159~MHz}=
37{\rm
~Jy}$ (Edge et al. 1959) and the optical jet counterpart was discovered by Crane
et al (1993) with the pre-COSTAR FOC instrument on HST. Since this initial
optical discovery, over $50{\rm ~ks}$ of HST observing time has been dedicated
to this source alone (e.g. Sparks et al. 1994; Baum et al. 1997, 1998; Hutchings
et al. 1998; Martel et al. 1999, 2000; Noel-Storr et al. 2003). More recent
observations have concentrated on deep optical polarimetry with WFPC2 (Perlman
et al. 2006) and optical and UV polarimetry with ACS (Capetti et al. 2007). It
was first detected in X-rays with {\it Einstein} (Elvis et al. 1981), 
and more recently observed by {\it
Chandra} and {\it XMM-Newton}, with the nuclear emissions studied by Donato et
al. (2004) and Evans et al. (2005), and the hot atmosphere by Sun et al. (2007).
Also, several radio
observations have been performed with the VLA (Lara et al. 1999; Lara et al.
2004; this paper), the global VLBI network (Lara et al. 1997) and the EVN and
MERLIN arrays (Baum et al. 1997; Lara et al. 1999; Lara et al. 2004). 

In this paper we attempt to bring all of this observational information into a
coherent picture of the physics of the 3C 264 jet. 
In \S 2 we detail the observations
and data reduction techniques used. In \S 3 we present comparisons between
optical and radio polarimetry, optical and radio spectral index maps, and a
comparison of the resolved and unresolved jet emissions from {\it Chandra}.
Lastly, in \S 4 we discuss the implications of these results, and present a
simple model that attempts to explain them. Throughout the paper we assume a
flat Friedmann cosmology with $H_0 = 72{\rm ~km ~s^{-1} ~Mpc^{-1}}$, 
(corresponding to the latest WMAP cosmology,
see Dunkley et al. 2009), giving a projected linear scale of $0.42{\rm ~kpc/''}$.

\section{Observations and Data Reduction}

We obtained polarimetric observations of 3C 264 with the WFPC2 instrument
aboard {\it HST} as part of the Cycle 10 observing program GO-9142. As part of
this same proposal we obtained 38.3 ks of {\it Chandra} observations of this
source.  We also
observed this object with the {\it VLA} in both A and B configurations at $8.5
{\rm ~GHz}$ and $22.5{\rm ~GHz}$ as part of observing program AP0439.  Further
observations in many wavebands and instrument configurations with {\it
HST} have been taken from the
online data archives (Table 1). What follows is a discussion of the
individual observing programs, as well as the methods used to reduce them.

\begin{deluxetable*}{llccccc}
\tablecolumns{7}
\tablewidth{0pt}
\tabletypesize{\small}
\tablecaption{{\it HST} Observations \label{HSTTable}}

\tablehead{
\colhead{Instrument} &
\colhead{Filter/Polarizer} &
\colhead{$\lambda_\circ^a$} &
\colhead{FWHM$^b$} &
\colhead{Num Exp$^c$} &
\colhead{Obs Date} &
\colhead{Time (s)$^d$}}

\startdata
WFPC2/PC      & F702W          & 6919  & 1382 & 2 & 12/24/1994 & 280  \\
WFPC2/PC      & F606W          & 5997  & 1502 & 1 & 04/16/1995 & 500  \\
WFPC2/PC      & F791W          & 7881  & 1224 & 2 & 05/19/1996 & 750  \\
WFPC2/PC      & F547M          & 5483  & 484  & 2 & 05/19/1996 & 900  \\
NICMOS/NIC2   & F110W          & 11285 & 3844 & 4 & 05/12/1998 & 448  \\
NICMOS/NIC2   & F160W          & 16030 & 2772 & 4 & 05/12/1998 & 448  \\
NICMOS/NIC2   & F205W          & 20714 & 4314 & 4 & 05/12/1998 & 896  \\
STIS/NUV-MAMA & F25QTZ         & 2354  & 995  & 1 & 02/12/2000 & 1800 \\
STIS/FUV-MAMA & 25MAMA         & 1374  & 318  & 1 & 02/12/2000 & 1705 \\
STIS/FUV-MAMA & F25SRF2        & 1469  & 281  & 1 & 02/12/2000 & 1700 \\
WFPC2/PC      & F702W          & 6919  & 1382 & 9 & 03/30/2002 & 1400 \\
\\
ACS/HRC       & F606W/POL0V    & 5895  & 1588 & 2 & 12/05/2002 & 300  \\
ACS/HRC       & F606W/POL60V   & 5895  & 1588 & 2 & 12/05/2002 & 300  \\
ACS/HRC       & F606W/POL120V  & 5895  & 1588 & 2 & 12/05/2002 & 300  \\
ACS/HRC       & F330W/POL0UV   & 3372  & 410  & 2 & 12/05/2002 & 300  \\
ACS/HRC       & F330W/POL60UV  & 3372  & 410  & 2 & 12/05/2002 & 300  \\
ACS/HRC       & F330W/POL120UV & 3372  & 410  & 2 & 12/05/2002 & 300  \\
WFPC2/PC      & F555W/POLQ$^e$ & 5439  & 1232 & 8 & 03/06/2003 & 6500 \\
WFPC2/PC      & F555W/POLQ$^e$ & 5439  & 1232 & 8 & 04/25/2003 & 6500 \\
WFPC2/PC      & F555W/POLQ$^e$ & 5439  & 1232 & 8 & 06/01/2003 & 6500 \\
\enddata

\tablenotetext{a}{Pivot wavelength of filter in \AA.}

\tablenotetext{b}{Approximate FWHM of filter bandpass in \AA.}

\tablenotetext{c}{Number of images in raw data set (including CR-SPLITs).}

\tablenotetext{d}{Total integration time of all images in seconds.}

\tablenotetext{e}{Due to non-rotatability of the WFPC2 polarizer (POLQ), {\it
HST} was allowed to precess between observations over a $\sim 2$ month time
period, simulating the nominal polarizer angles of $0^\circ$, $60^\circ$, and
$120^\circ$. See text for details.}

\end{deluxetable*}

\subsection{{\it HST}} 

We observed 3C 264 in three separate epochs with {\it
HST}'s WFPC2 instrument during an approximately two month time period between
March and June of 2003 (see Table \ref{HSTTable}). We used the
wideband F555W filter, combined with the POLQ polarizer. The time separation for
these observations was necessary {\footnote{To obtain high-quality polarimetry,
it is necessary to take observations through three polarizers, with the second
and third nominally
oriented at either 60$^\circ$ and 120$^\circ$ or 45$^\circ$ and 90$^\circ$ with 
respect to the first}} since we wanted the object to reside on the PC
chip, and the POLQ filter rotates only through $\sim50^\circ$. All of the 
rotated
positions leave the PC either severely vingetted or not covered by the same
polarizer. Thus, {\it HST} was allowed to precess between observations so that
the polarizer angles approximated the standard $0^\circ$, $60^\circ$ and
$120^\circ$ polarizer angles. Windows of $15^\circ$ in orientation were allowed
for scheduling purposes, and to avoid the jet lying along a diffraction spike,
ORIENTs were chosen carefully. The PC chip was chosen due to its higher
angular resolution as compared to
the WF chips. Subpixel dithering was used in order to fully
sample the PSF, and hence achieve diffraction limited {\it HST} resolution in
this waveband.

{\it HST} ACS polarimetry was also obtained for 3C 264 as part of GO program
9493, in December of 2002 (Table \ref{HSTTable}). These observations were taken
with the {\it High Resolution Channel} (HRC) of the ACS, combined with the
wideband optical F606W filter, and the F330W UV filter. The optical polarimetry
was taken through the three visible polarizers aboard ACS, namely POL0V, POL60V
and POL120V, and the UV polarimetry was obtained using the three UV polarizers,
namely POL0UV, POL60UV and POL120UV. The ACS is better suited to
polarimetry than WFPC2, in that it has higher sensitivity - more than seven
times that of WFPC2 at $\sim 6000$ \AA\ \citep{Heyer04} - it has three
polarizers oriented very near to the ideal polarizer angles, and the point
spread function is better behaved across the chip. The fact that ACS has three
properly oriented polarizers (for both optical and UV) allows the neccessary
polarization observations to be taken in the same epoch.
This simplifies the data reduction process significantly, by
easing the image registration process with un-rotated PSFs, and eliminating the
effects of varying PSFs across the imaging chips, both of which complicated the
WFPC2 polarimetry reduction (see Perlman et al. 2006 for details). 
It also eliminates the possible complication of jet variability.  However, 
the exposure times for these two data sets are much shorter than our WFPC2
polarimetry, hence we are not able to detect jet emission to the same levels,
despite the increased sensitivity of ACS.

We also used the online {\it HST} data
archive to obtain imaging observations of 3C 264 in numerous wavebands (see
Table \ref{HSTTable}), allowing us to explore the optical spectrum of the jet.
The results cover eight years of observations,
at wavelengths from
$\sim1400${\rm~\AA} to $2{\rm ~\mu m}$, and include data from WFPC2, 
STIS as well as NICMOS.  

All of the WFPC2 data were reduced using standard techniques with {\sc stsdas}
in {\sc iraf}. {\footnote{For an introduction to {\sc iraf} and {\sc stsdas},
see http://iraf.net and
http://www.stsci.edu/resources/software\_/hardware/stsdas}}  The images were calibrated using the best dark count images and
flat field images available.  Cosmic-ray rejection was done using the {\sc
stsdas} task {\sc crrej} for all but one data set. The F606W image was a single,
non-CR-SPLIT observation, and as such, could not have the cosmic rays removed. 
Fortunately, there were no cosmic ray events in the jet region, so the cosmic
rays were simply masked during galaxy subtraction (see below). The resulting
cosmic-ray free images were then {\it drizzled} on to the native WFPC2/PC pixel
scale of $0.0455''/{\rm pixel}$ and converted to counts/s, using the {\sc
stsdas} task {\sc drizzle}. This task corrects for the field distortion of
WFPC2, for which we used the latest geometric correction coefficients for the PC
chip \citep{Anderson03}. 

The above steps were also taken for each individual observation through the
polarizer, except that these data were sub-sampled onto half the native PC
pixel scale - $0.02275''/{\rm pixel}$.  Since our polarimetry observations are
quite deep (four orbits in all; nearly two hours through each polarizer), we
were able to 
 detect jet emission out to nearly $2''$ from
the core.  As we used sub-pixel dithering, we were able to fully sample the
PSF on  the PC chip and achieve diffraction limited resolution  
\citep{Heyer04}. Similarly to the WFPC2 data, the
ACS/HRC data listed in Table \ref{HSTTable} were reprocessed with the best dark
count and flat field images available in the archive, using the procedure 
outlined in Pavlovsky (2005). These data were further
reduced using the PyRAF task {\sc multidrizzle}.  In our case, since there is
only a single {\sc cr-split} observation for each polarizer, this multi-faceted
task was limited to cosmic ray rejection and geometric distortion correction.
For the latter we used the latest geometric distortion coefficients for the HRC
\citep{Anderson04}.

The STIS NUV- and FUV-MAMA data listed in Table \ref{HSTTable} were calibrated
using STScI's On-The-Fly-Reprocessing (OTFR) which uses the best available flat
field images and dark count corrections. OTFR also {\it drizzled} the images
onto the native FUV-MAMA and NUV-MAMA pixel scales of $0.0246 \times
0.0247''/{\rm pixel}$ and $0.0245 \times 0.0248''/{\rm pixel}$ respectively.
The sky background for each image was estimated to be simply the mode of all
pixels, using the {\sc iraf} task {\sc imstatistics} with 2 iterations of
$3\sigma$ rejection.  This approach was necessary since there was
only a single exposure for each detector. Galaxy subtraction (see below) was
only performed for the NUV-MAMA image, since galaxy light is minimal, if
present, in the FUV-MAMA images.  The STIS data were the only HST observations
for which sky subtraction was possible, given the large angular size of the 
host galaxy, which filled the field of the WFPC2, ACS/HRC and NICMOS data.  

The NICMOS data listed in Table \ref{HSTTable} were taken with the NIC2
camera, in {\sc multiaccum} mode. These data were also reprocessed using OTFR,
with the best available flat fields and dark count images. For the F205W data
the pipeline flats did not completely account for the NIC2 response at this
wavelength and for these particular observations. The spatial scale of the
variations is however larger than the extent of the region of interest (i.e.
the jet), and we were able to subtract the galaxy to achieve a residual noise
level of only $0.026{\rm ~counts ~s^{-1} ~pixel^{-1}}$. The ``pedestal'' bias
\citep{Dickenson04} was removed using the {\sc stsdas} task {\sc pedsub} (based
on the original code, ``unpedestal'',  by R. van der Marel), which is designed
to handle crowded fields. This produced a better reduction of the quadrant bias
than did the other commonly used task {\sc pedsky}. For each of the three
filters there were four dithered observations, which were {\it drizzled}
separately onto a common frame with the nominal NIC2 pixel scale of
$0.0775''/{\rm pixel}$. The individually {\it drizzled} images were then median
combined using the {\sc iraf} task {\sc imcombine}, rejecting the lowest of the
four values at each pixel. This was done to account for the fact that these
observations unfortunately placed the jet across the NIC2 erratic middle column
\citep{Dickenson04}, which shows up as a dark strip.  Since {\sc drizzle} is
more or less a shift-and-add algorithm, it does not do well with artifacts of
this nature. Hence, rejecting the lowest valued pixels seemed the best
solution.

Once the calibrated images were made, an isophotal model of the host galaxy
was obtained using the {\sc iraf} tasks {\sc ellipse}
and {\sc bmodel} \citep{Perlman99}. The jet and globular clusters were masked
out during the isophotal fitting, as their inclusion 
leads to oversubtracted ring-like artifacts in the
subtracted images. In addition to this masking, outlying pixels were excluded
using a $3\sigma$ rejection in the isophotal fits.  
The resulting isophotal fit is shown
in Figure 1 for the WFPC2 polarization image, 
compared to the radial profile extracted from the polarization 
data.  As can be seen, the galaxy subtraction
process was made difficult by the nearly circular dust ''disk'' extending out
to $\sim 0.8''-0.9''$ from the galaxy core (Martel et al. 1998, 2000), 
as well as a ring of enhanced
emission just outside of this dust disk.  To overcome this,
we interpolated our models linearly over most of the dusty region,
including only those isophotes that were well centered on the galaxy core, and
very close to circular. The latter criterion was chosen since the galaxy itself
is classified as an E0 elliptical, and hence nearly all of its isophotes have
ellipticities $\lsim 0.1$.

\begin{figure}
\centerline{\includegraphics*[width=3.5in]{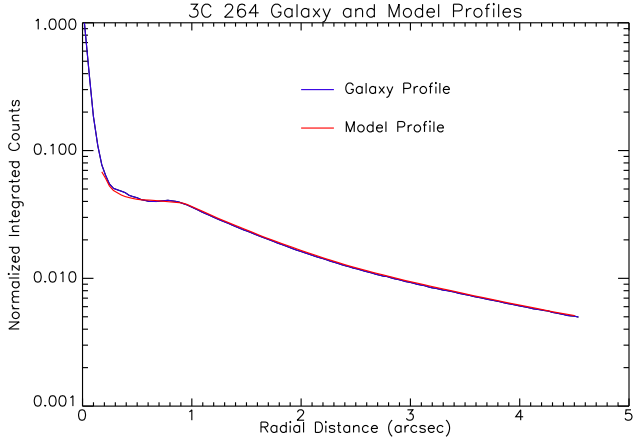}}
\label{ProfilesFig}
\caption{Integrated radial profile of a single polarizer image from WFPC2
plotted with the radial profile of our best fit isophotal model. The jet
emission can be seen between $0.3''$ and $1.0''$, as slight enhancements in the
galaxy profile. Note that the galaxy model does not extend inwards of
$\sim0.2''$ since the core of the WFPC2/PC PSF is not well modelled by
ellipses.}
\end{figure}

After the galaxy light was subtracted, the polarization images were combined to
produce Stokes I, Q and U images, as well as fractional polarization (\%P) and
polarization position angle (PA) images. For the details of this process see
Perlman et al (2006).  The resulting optical polarimetry is shown in Figure 2.
For comparison, radio imaging and polarimetry (next section) are shown in 
Figure 3.

\begin{figure*}
\centerline{\includegraphics*[width=6in]{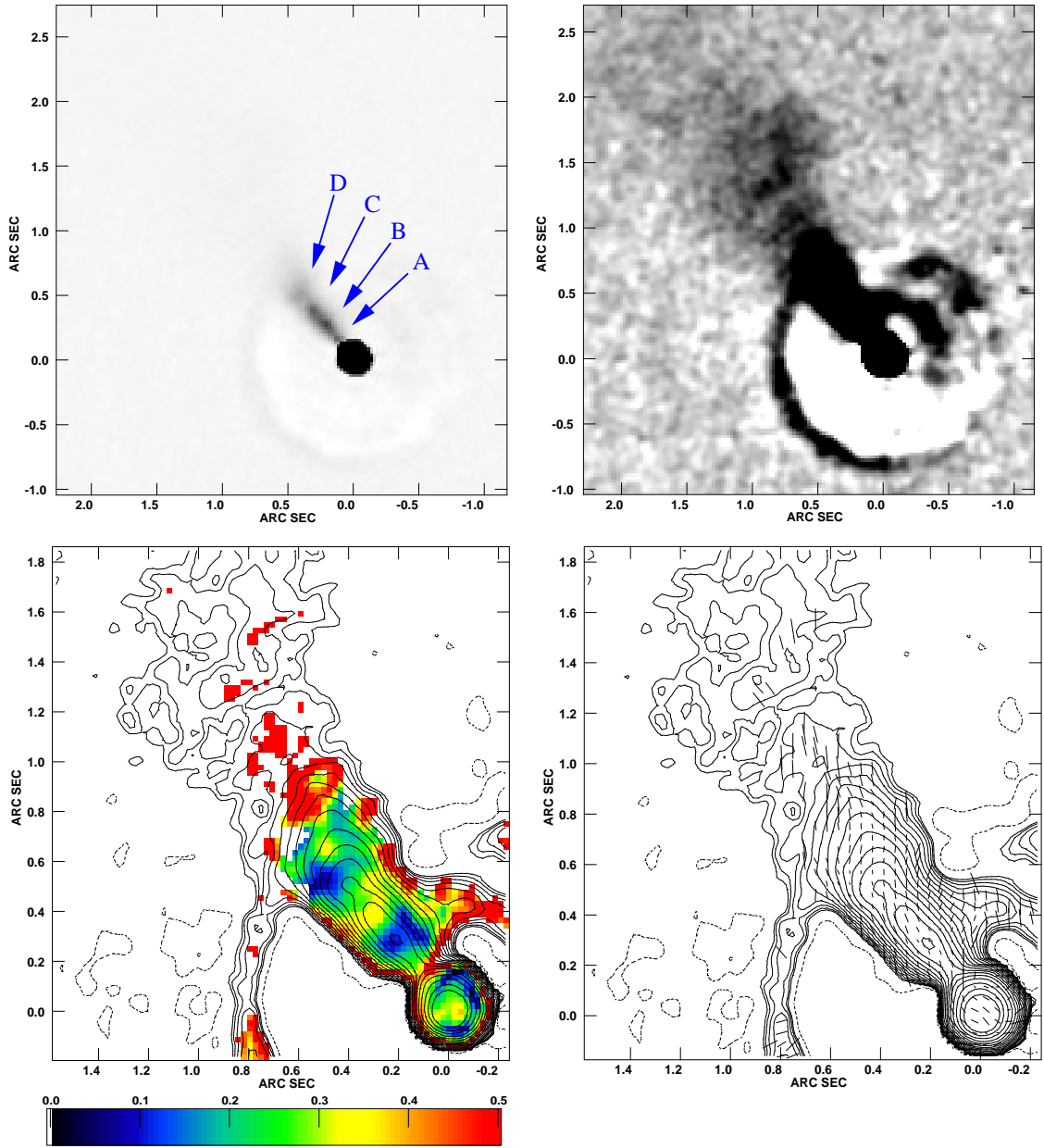}}
\label{OptPolFig}
\caption{Optical (HST F555W) polarimetry of the 3C 264 jet (copied with 
permission from
Perlman et al. 2006). The image scale is 440 {\rm pc/arcsec}.  {\it Top Left}:
Stokes I image scaled to bring out fine-scale structure within the knots. {\it
Top Right}: Stokes I image with stretch that saturates on bright regions, but
helps bring out faint structure in this jet.  In this view the jet can be seen
beyond the dust ring at $1''$ from the nucleus (Sparks et al. 2000). {\it
Bottom Left}: Fractional polarization (colors) overlaid with Stokes I
contours - red indicates $\gsim 45\%$.  {\it Bottom Right}: Stokes I contours
with polarization ({\it B} field) vectors; a vector $0.1''$ long represents
150\% polarization; and contours are spaced by $\sqrt{2}$, with a base contour
level of 1.783 ct/s.}
\end{figure*}

\begin{figure*}
\centerline{\includegraphics*[width=5in]{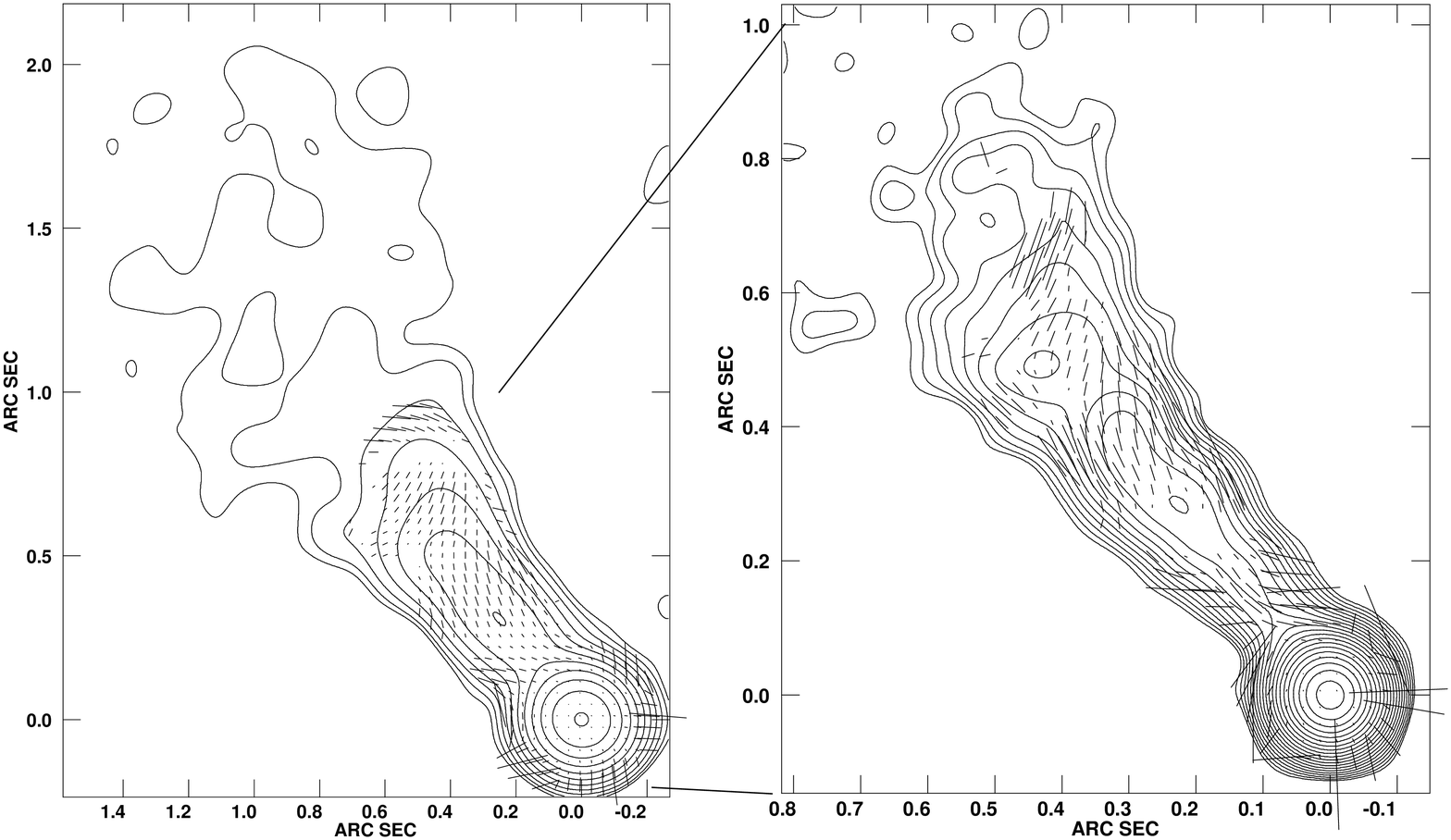}}
\label{RadPolFig}
\caption{Fractional polarization maps made with VLA A-array data at 22.5 GHz. 
Contours represent Stokes I,
and are spaced by powers of 2. Vectors are proportional to
fractional polarization -- $0.1''$ corresponds to $100\%$ polarized flux -- and
are rotated to represent projected magnetic field direction. {\it Left}: Image
made with a $800{\rm k\lambda}$ taper applied.  The base contour
level is 0.138 mJy/beam.  This taper reveals
the faint part of the jet beyond $1''$ from the core, which is also detected in
the optical. {\it Right}: The same data, mapped without a UV taper.  The
base contour level is 0.171 mJy/beam.  This
full resolution map clearly shows the widening jet (Lara et al. 2004; see also 
Figure 4)
as well as the dramatic drop in signal outside the ''ring''. Also note how the
B-field vectors follow the flux contours until just before knot D, where they
turn to the north-west, nearly perpendicular to the jet. This may be evidence
of a weak shock.}
\end{figure*}

The imaging observations (including the Stokes I images)
were resampled onto the WFPC2/PC native pixel scale of $0.0455''/{\rm pixel}$,
and registered to one another by fitting elliptical gaussians to the galaxy
core. The registration was done in this manner since, due to color gradients
and the lack of UV emission from both the galaxy and globular clusters, the
images do not lend themselves to good cross-correlation across many wavebands.
We converted all images to flux density units (millijansky) using the STSDAS
SYNPHOT package.  We convolved all images to the resolution of the NICMOS F205W
data by convolving them with circular Gaussians, following the method of 
Perlman et al. (2001).  
The resulting optical spectral index map is shown in Figure
4; note that we use the
convention $S_\nu \propto \nu^{-\alpha}$.

\begin{figure}
\label{OptSpixPol}
\centerline{\includegraphics*[width=3.5 in]{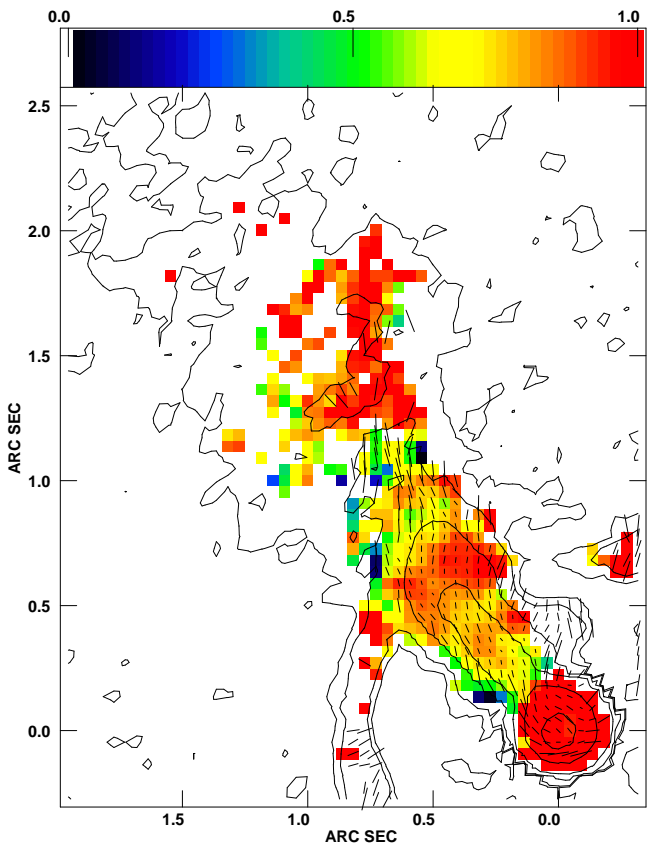}}
\caption{Optical fractional polarization map, obtained using the F555W WFPC 
data, overlaid on optical spectral index
map. Base contour level is 6.6$\times 10^{-4}$ ct/s/pix, in optical Stokes I,
with contours spaced by factors of 4. Note that this image used the 
native PC pixel scale of 0.0455''/pix, while Fig. 2 was sub-samled by 2.  
Vectors
represent projected B-field direction, and their length is proportional to \%P
- a vector $0.1''$ long represents 150\% polarization.  The dust ring can be 
seen at approximately $0.8''$ from the core.  Note the structure discussed in
\S \S 3.1 and 4.1.}
\end{figure}

\subsection{{\it VLA}}

We observed 3C 264 with the {\it VLA} in K-band ($22.5{\rm ~GHz}$) and X-band
($8.5{\rm ~GHz}$) in B configuration in June 2002, and in A configuration in
August 2003 as part of program AP439. 
For these observations 3C 286 was used as the primary flux density calibrator
source, and the QSO J11503+24179 as the primary phase calibrator and secondary
flux density calibrator. The same two 
sources were also used to determine the instrumental
polarization parameters.
The K-band observations were done using the VLA's
fast-switching mode to calibrate out any phase fluctuations due to
tropospheric variations. This entailed alternating between $125{\rm ~s}$
scans of 3C 264 and $45{\rm ~s}$ scans of the phase calibrator. The primary
flux density calibrator was observed with $70{\rm ~s}$ scans every hour. The
X-band observations were used not only for imaging, but also as a pointing
reference for the high-frequency observations. Two IFs were used in each
band, each with a bandwidth of $50{\rm ~MHz}$, and the average of the two was
used to produce the final maps. In addition to our observations, we have also
obtained C-band ($5{\rm ~GHz}$) A-array data from the online {\it VLA} archive
(see Lara et al. 2004), in order to make a matched resolution
$5{\rm ~GHz}$ to $8{\rm ~GHz}$ spectral index map. For this data set also, 
3C 286
and 3C 48 were used as the primary flux density calibrators, and the
QSO J11503+24179 was used as the phase calibrator and secondary flux density
calibrator. 

The calibration of the C-band and X-band datasets was straightforward,
following closely the recommended procedure in Chapter 4 of the {\sc AIPS}
Cookbook.  Very little flagging was required for these data sets, and all of
the antennas in the array were well behaved. Once calibrated, a clean map was
made with several rounds of self-calibration, and deconvolution using the {\sc
AIPS} tasks {\sc clean} and {\sc calib}. For the deconvolution we used the
``robust'' or Briggs' weighting scheme to weight the visibilities and
a restoring clean beam of $0.482'' \times 0.365''$ for the C-band data, one of
$0.246'' \times 0.218''$ for the X-band A-array data, and one of $0.76'' \times
0.68''$ for the X-band B-array data. The rms noise level achieved in the C-band
Stokes I image is $65{\rm ~\mu Jy ~beam^{-1}}$ - on the order of the
theoretical noise limit of $46{\rm ~\mu Jy ~beam^{-1}}$. The achieved noise
levels in the X-band data were similarly close to the thermal noise limit. The
other Stokes parameters were not solved for in these data sets, as the C-band
polarization structure has already been published (see again Lara et al 2004),
and our X-band data lacks polarization calibrator observations.

The reduction of the K-band data was performed using mostly
standard techniques in {\sc aips}, following closely the recommended procedure
given in Appendix D of the {\sc aips} cookbook {\footnote{See
http://www.nrao.edu/aips/cook.html}}.  The first $10{\rm ~s}$ of every
scan were flagged to account for mispointings at the start of scans.  This was
particularly important for this fast-switched data since the array elements
have their pointings changed very frequently. Antenna 1 was flagged in all of
the data due to very low amplitude measurements with respect to the rest of the
array. This was in addition to any antennas that were flagged by the on-line
data quality monitors. Also in both the A-array and B-array data, there was a
tropospheric noise component, forcing us to flag the first and last hour of each
data set. 

After the initial flagging was done, we first solved only for the phase
solutions of our primary phase calibrator, assuming a point source model.  This
allowed us to calibrate out any phase fluctuations induced by the troposphere.
These solutions were found in $10{\rm ~s}$ intervals, and applied to only the
phase calibrator and 3C 264. Amplitude and phase solutions were then found for
both the primary flux density calibrator, 3C 286, as well as the secondary flux
calibrator. For the former, a clean component model provided by {\sc aips} was
used, and for the latter, a point source model. The solutions for 3C 286 were
used only to bootstrap the flux density scale for the other objects - phase
solutions based on 3C 286 were not applied to other objects. Polarization
calibration was performed using QSO J11503+24179 to solve for the polarization
leakage terms (the ``D-terms''), and using 3C 286 as the polarization position
angle calibrator. Once the calibration was complete, clean maps were made as
above, using several rounds of self-calibration and deconvolution with Briggs'
weighting. A restoring clean beam of $0.0844'' \times 0.0826''$ was used for the
A-array data and one of $0.264'' \times 0.254''$ for the B-array data. Again the
rms noise limit achieved in each configuration was on the order of the
theoretical noise limit - $57{\rm ~\mu Jy ~beam^{-1}}$ for the A-array Stokes I
image and $68{\rm ~\mu Jy ~beam^{-1}}$ for the B-array, compared to the
theoretical limit of $42{\rm ~\mu Jy ~beam^{-1}}$ for both configurations. For
these data, we also mapped the Stokes Q and U, using the same ``clean boxes'' as
were used to make the Stokes I image. From these the \%P and PA maps were made,
with the well-known Rician bias in P \citep{Serkowski62} removed using the {\sc
AIPS} task {\sc polco}.  The resulting images are shown in  Figure 3, and
discussed in \S 3.  Radio and radio-optical spectral index maps are shown 
in Figure 5.

\begin{figure*}
\centerline{\includegraphics*[width=6.5in]{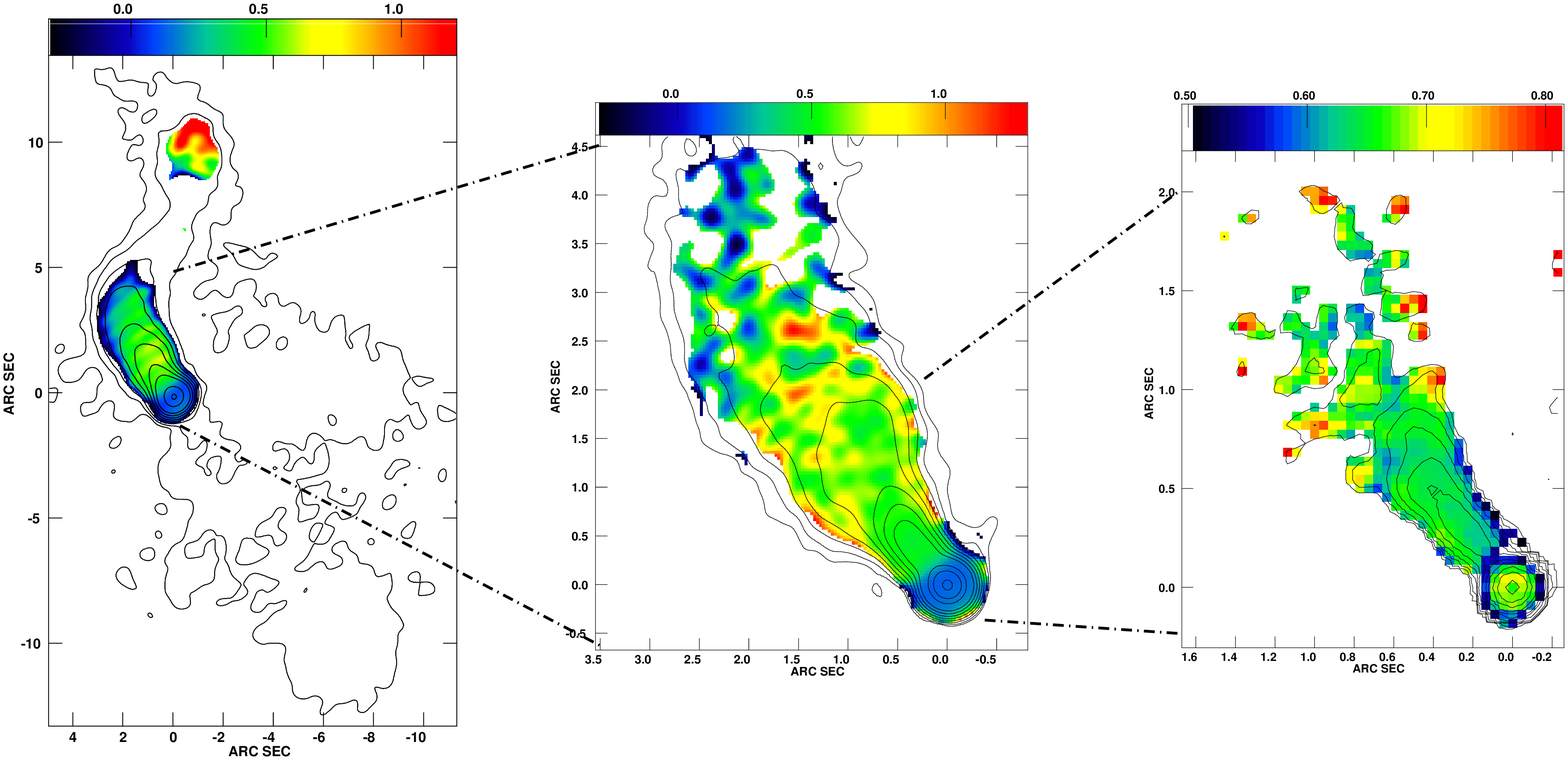}}
\label{SpixFig}
\caption{Radio and Radio-optical spectral index maps.  In all plots, the 
contours are spaced by powers of 2 and the colour table is as given.  
{\it Left}: Spectral index map between $5.0{\rm ~GHz}$ and 
$8.5{\rm ~GHz}$, at $0.5''$ resolution. Base contour level 
is 0.203 mJy/beam. {\it Middle}: 
Spectral index map between $8.5{\rm GHz}$ and $22.5{\rm GHz}$,
at $0.3''$ resolution.  Base contour level is 0.835 mJy/beam.
{\it Right}: Spectral index
map between $22.5{\rm GHz}$ and F555W HST data, at $0.2''$ resolution. 
Base contour level is 0.013 mJy/pix.  Note 
the relative constancy of the spectral indices in the inner $1''$ of the 
jet.}
\end{figure*}

\subsection{{\it Chandra} Observations} 

We observed 3C 264 with the {\it Chandra X-ray Observatory
(Chandra)} on January 24, 2004, using the ACIS-S instrument
detector in {\sc faint} mode with the object centered on the
back-illuminated CCD, S3. This CCD was chosen due to its
sensitivity in the soft X-ray regime. We used a $1/8$
sub-array of the chip to mitigate photon pileup.  A total of 38.3 ks was 
granted to this program, with 34.76 ks remaining after flagging
out background flares.

The imaging data were processed using CIAO version $3.4.0$, and CALDB
version $3.2.2${\footnote{For a discussion of CIAO see
http://cxc.harvard.edu/ciao.}}.  A new level 2 events file was created
following the recommended procedure in the CIAO science
threads, after which pixel randomization and the readout
streak were removed. The sub-pixel event repositioning (SER)
algorithm described by Li et al.\ (2003) was then applied to
the event data, and an image was generated from the
resulting SER-processed events by resampling at $0.246''$,
i.e., one-half of the native ACIS-S pixel scale. 
Following SER, we also applied a 
maximum-likelihood deconvolution algorithm to the imaging data.  
This allowed us to maximize the angular resolution of the data and 
distinguish the jet from the core. The resulting image is shown in
Figure 6 as contours, with the F555W image overplotted in color.

While CIAO version 3.4 is adequate for our imaging analysis, 
more recent available calibrations (CIAO version 4.1.2 and 
CALDB version 4.1.3) were used to make a new level 2 events file
and construct calibration data for spectral fitting.
A spectrum of the X-ray source core was extracted from the
events file using a circular aperture with a radius of
$1.23''$ (5 post-SER pixels) and a local background region, via the CIAO
script {\sc psextract}.    For AGN this
region typically contains $95\%$ of the core flux (Marshall
et al 2005), and so no aperture correction was applied.  The instrument
response matrix and ancilliary responses were generated using the
CIAO tasks {\sc
mkacisrmf} and {\sc mkarf}. We discuss in \S\S 3.3,4 the X-ray 
and broadband spectrum of the jet and its physical implications.

\section{Results}

In this section, we present the results of the above analysis, and compare them
to previous work. We start with general morphological descriptions of the 3C
264 jet in all analyzed wavebands, as well as the broadband spectral
characteristics. Then we discuss the HST optical and VLA radio polarimetric
properties of this source, as well as its X-ray morphology and spectrum.

\subsection{Radio-Optical Morphology and Spectrum}

3C 264 has a Fanaroff-Riley type I radio structure (e.g., Fanaroff \& Riley
1974) with a head-tail morphology and bent jet.  As can be seen in our 
radio maps (Figure 5), the radio jet extends $\sim 25''$
along a direction that is initially northeast, but bends through $\sim 90^\circ$ to
the west at $3''-6''$ from the nucleus, and then back towards the northeast at 
greater distances (Gavazzi et al. 1981; Bridle \& Vallee 1981; Baum et al. 1988; 
Lara et al. 1999).  On these size scales, the jet is seen only in the radio.
Our
$8.5 {\rm ~GHz}$, B-array map  (left panel of
Figure 5) shows a morphology similar to that described in Lara
et al. (1997, 2004) at lower frequencies.  All these images show a relatively
smooth morphology between 2 and 4 arcseconds from the core, a region where the
jet begins a bend through about 90 degrees.  By a distance of 4 arcseconds from
the core, the jet has reached a nearly north-south orientation, and then 
becomes northwesterly for several arcseconds before bending back to the 
northeast at about 10 arcseconds from the core at the location of a fairly
significant knot. Radio structure on larger scales (after this second bend) has
been discussed by Lara et al. (1997, 2004).  

Along with the radio maps, Figure 5 also shows the radio spectral indices.  The
left-hand panel shows the radio spectral index between 5 and 8.4 GHz.  As can be
seen, the spectral index $\alpha_5^{8.4}$ is $\sim 0.5$ throughout the jet's
first $\sim 4$ arcseconds, out to the location of the first bend.  The 
structure at the locus of the second bend is somewhat steeper spectrum, 
particularly at its downstream end.
The middle
panel of Figure 5 shows the spectral index of the jet between 8.4 and 22.5 GHz. 
As can be seen, we see a spectral index of $\alpha_{8.4}^{22}  \sim 0.5$
throughout the entire jet out to $\sim 2''$ from the core, with somewhat
flatter  spectral indices at greater distances, where the bend begins. 
From the C-X spectral index  map, we find the radio spectral
index of the core of 3C 264 to be $\alpha_{core} \simeq 0.14$ , consistent with previous results
(cf. Lara et al. 2004).  The radio spectral index of the inner jet is also
consistent with previous results, around $\alpha \sim 0.5$. 

The optical morphology of the inner $1''$ of the 3C 264 jet 
(Figure 2) is similar to that seen in the 22.5 GHz A-array map 
(Figure 3). Both maps show a jet that has a surface brightness
which varies smoothly along the ridgeline.
A few faint knots appear, and we have termed them
A-D in order of increasing distance from the nucleus (pointed out in Figure 2). 
The jet fades precipitously outside of $\sim 1''$ from the core (this last section 
is not 
seen well in the UV image), where a prominent dusty disk or "ring" is located.
While this might suggest an
interaction between the disk and the jet, the jet viewing angle has been found
to be $\sim 50^\circ$ (Baum et al. 1997; Lara et al. 1997).  The disk is
more likely viewed face-on, however, due to its nearly
circular shape and apparent lack of rotation (Baum et al.
1990, 1992).  Thus the likelihood of interaction is small, 
although not out of the question (see Baum et al. 1997 for a
discussion). Outside of this ``disk'', the F555W Stokes I image  
shows jet emission continuing out to roughly $2''-2.5''$ from the
core, emission that is also seen in both our 22.5 and 8.4 GHz data (Figure 3 
and also middle panel of Figure 5) as well as the MERLIN 1.5 GHz 
map of Lara et al. (2004).

We also produced a map of the radio-optical spectral index,  
shown in Figure 5's right-hand panel.
This varies between $\alpha_{RO}=0.6-0.7$ over the vast majority of the inner
jet (i.e., within $1''$ of the core).  The radio-optical spectral indices seen
along the jet centerline are slightly flatter (i.e., harder) than those seen
along the jet edges (throughout the optically seen portion of the jet), as  well
as possibly along the jet's northern edge at distances $<0.5''$ from the
nucleus.  The $\alpha_{8.4}^{22}$ map also shows steeper spectral indices along
the jet edges than in the center, beginning at distances of $\sim 0.5''$ from
the nucleus, as well as the slight flattening of the spectrum at the jet's
northern edge within the first $0.5''$. Thus, contrary to some other jets (e.g.,
M87, Perlman et al. 2001), the first 2 arcseconds of the 3C 264 jet (i.e., the
portion seen in the optical) shows a remarkable consistency in its radio-optical
spectral energy distribution. 

\begin{figure}
\centerline{\includegraphics*[width=3.5in]{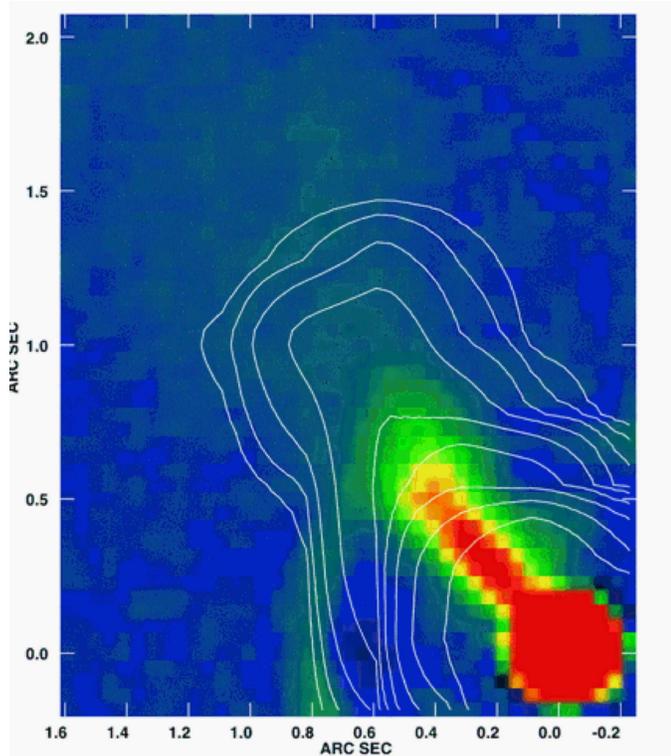}}
\label{chfig}
\caption{The Chandra X-ray image of 3C 264 (contours), 
plotted over the F555W Stokes I image (color).  The {\it Chandra} data have
been processed according to the procedure in \S 2.3, deconvolved 
using a maximum entropy algorithm, interpolated onto the PC pixel scale,  
and then smoothed with a 1-pixel Gaussian.  X-ray contours are spaced by factors 
of 2, with a base level of 5.725$\times 10^{-2}$ cts/pix.
Note that X-ray emission is detected from at least two distinct regions in 
the jet.}
\end{figure}

 

The HST data show that 
the morphology of the jet is smooth and consistent from the IR to the
UV, at least at the 0.1$''$ level.   This is reflected in the smooth 
variations seen in the optical spectral index map of Figure 4. 
%
But at the smallest scales we can resolve  
(as seen on both HST UV images as well as the full-resolution F555W Stokes I
image), the knot-like surface brightness enhancements A-D 
become pronounced, indicating that the overall smoothness 
may be simply a resolution issue. This is supported by the more
knot-like structure seen in the $1.6{\rm ~GHz}$ MERLIN maps of Lara et al
(2004). 

In Figure 7, we plot the optical  spectrum and spectral energy distribution for
the  jet and core, as seen in our data.    The optical spectral index for the
entire jet (all pixels $>3\sigma$), is found to be $\alpha_O = 0.77 \pm 0.03$.
This is harder than that of the core, 
in contrast to previous results around $\alpha_O \sim 1.4$ of Crane et al.
(1993) and Lara et al. (2004). These latter points are also plotted in the top
left-hand panel of Figure 7.  As can be seen, the fluxes
for the two previous studies are not consistent with one another:  at a
similar frequency,  Crane et al. (1993) found a flux nearly a factor two higher
than did Lara et al. (2004).  However, Crane et al. 
did not model  the galaxy, instead using 20-pixel-wide strips located above
and below the jet to  subtract out galaxy emission.  The likely result of their
procedure is to  undersubtract the galaxy, particularly in the red and indeed,
we can see that   their F342W point falls within the error bars of (although
slightly above) our  F330W flux, while their F502W flux point is nearly a factor
two higher.  This  corroborates our assertion that their spectral index estimate
was contaminated by incorrect subtraction of galaxy light.  For the Lara et al.
(2004) points,  the issue is more subtle, as we see fluxes very consistent with
the ones we report. However, as shown in Figure 6,  Lara et al. (2004)'s data
spanned only   $\sim 0.2$ decades in frequency, giving a small pivot in their
spectral index fits. Our data span more than a decade in frequency, which
removes ambiguity in the fit. In the same figure, we show lines  for best-fit
spectral indices in the infrared, optical and UV bands.  As expected, the errors
on those  fits are far larger than when all the HST data are used (by factors of
4-14); this illustrates the danger in using a spectral index based on  data
spanning only a small fraction of a decade in frequency.  Our data show no
evidence of a change in spectral index from the infrared to the optical;
however, there is possible evidence (2.3$\sigma$) for a steepening of the
spectral index to the ultraviolet.  The same plot shows the spectrum of the core
in the optical. As can be seen, the spectral index of the core throughout the
infrared to UV  is remarkably similar to that for the jet (although the core is
more than a factor two brighter), although the core shows no evidence for
hardening towards the ultraviolet.  This strongly suggests that the same
emission  mechanism dominates the emissions of the two, although other data 
also show a jet spectrum that is noticeably harder than that of the core in the 
blue (Hutchings et al. 1998).

\begin{figure*}
\label{SED}
\centerline{\includegraphics*[width=6in]{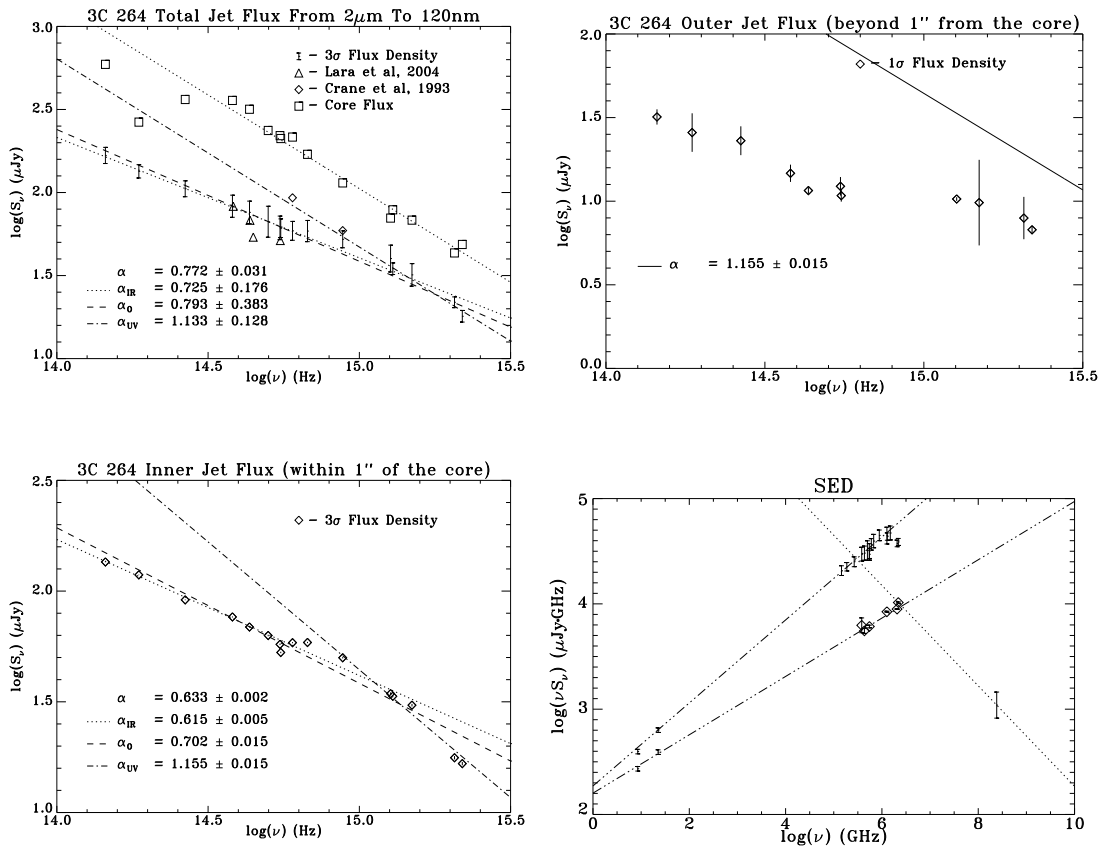}}
\caption{Optical Spectrum and Spectral Energy Distribution of the 3C 264 jet.
{\it Top Left Panel:}  Total Jet Flux and Spectrum in the Optical.  For the jet,
we show fluxes
from three studies here:  Our own (error ranges), Lara et al. (2004), and Crane et 
al. (1993).  We also show fluxes for the core as taken from our data.  For the
total jet flux, note the excellent overlap of our points with those from Lara et
al. (2004).  The flux points of Crane et al. (1993) are, however, clearly seen 
to be discrepant, particularly in the red.  See \S 3.1 for details.  {\it Bottom 
Left  Panel:}  Inner Jet Flux and Spectrum in the Optical.  
This plot shows fluxes from our HST 
data for the region of the jet within $1''$ of the core.  Lines are shown describing
the spectral indices in the radio-optical, IR, optical and ultraviolet.  
{\it Top Right Panel:}  Outer Jet Flux and Spectrum in the Optical.  This plot shows
fluxes form our HST data for the region of the jet at distances greater than $1''$
from the core.  
{\it Bottom Right Panel:}  Spectral energy distributions of the 
inner jet (upper values) and
outer jet (lower values). The dashed-dot lines represent a simple power-law fit
to the radio and optical data fot both the inner and outer jet, and the dotted
line represents the same fit to the UV to X-ray data for the outer jet. For the
inner jet, optical data with frequency above $10^{14.5}{\rm ~Hz}$ were not
included. For the outer jet, all optical data were included.  Lines are shown
describing the spectral fits.  See \S 3.1 for details.}
\end{figure*}

Figure 6 ({\it Bottom Left Panel}) shows the fluxes for the inner jet (within
$1''$ of the nucleus) along with spectral index fits for the entire HST band,
infrared, optical and ultraviolet.  As can be seen, the spectrum of the inner
arcsecond of the 3C 264 jet is somewhat flatter than that of the entire jet,
with an average spectral index $\alpha = 0.633 \pm 0.002$.  The spectrum of the
inner jet is slightly but significantly steeper (0.702 $\pm$ 0.015 compared to 
0.615 $\pm$ 0.005) in the optical than it is in the near-infrared, with a more 
significant steepening in the ultraviolet ($\alpha_{UV} = 1.155 \pm 0015$). 
These values are, it should be noted, within 1$\sigma$ of the values seen for
the jet as a whole, as is sensible, since the errors should be largest in the
outer jet, where the jet is faintest.  Those pixels therefore dominate the
larger errors in the  previous spectral index fits.    Figure 6 ({\it Top Right
Panel}) shows the spectrum of the outer jet (at distances greater than $1''$
from the core).   As expected, the errors are much larger for this region.  

Figure 4 shows a detailed
spectral index map for the optical regime, overplotted on the F555W
polarimetry.  This map shows significantly more structure than that seen in 
the $\alpha_{RO}$ map (Figure 5), indicating that the SEDs of some parts of the
jet are turning over in the optical.
The most significant feature of the optical spectral
index map is a softening by $\Delta \alpha_o \approx 0.2$ in a 0.2''
long region, nearly coincident with the location of the dust
ring.  This softening may be a result of galaxy subtraction residuals, or
it could be indicative of increased energy losses.  The latter would
suggest an interaction between the jet and the ring (as mentioned above). 
However, it is hard to comment on this definitively  
since the galaxy had to be modelled independently in each HST image as 
its inner $1''$ changed significantly from the IR to the UV. Outside
of the ring, the optical spectral index is generally steeper than that
seen in the inner jet, a feature that was also seen in the $\alpha_{RO}$ map.
Some interesting structure is also seen interior to the ring, 
where along the jet ridgeline we see flatter optical spectral near the loci
of the knots, with steeper values of $alpha_O$  both along the jet's
edges as well as in the region between knots B and C.  
We find the core's optical spectral index to
be $\alpha \simeq 0.9$, which, when the radio core spectral index is
taken into account, indicates a spectral break of $\Delta \alpha \sim
1$ in the core spectrum around $10^{13}{\rm ~Hz}$.

\subsection{Multi-band Polarization}

The optical polarization characteristics of the 3C 264 jet have been 
discussed already (using the same WFPC2 data) by Perlman et al. (2006).
However, it is useful to reprise the discussion here in view of the new 
VLA and ACS polarization data. 
We see an average fractional polarization level of $\sim 20\%$ along the
jet, roughly similar to that seen in the radio, and projected magnetic 
field vectors that generally follow the flux contours.  As
noted there, in our WFPC2 F555W data (Figure 2) we find a 
polarization morphology that is quite homogeneous,
in line with the overall smooth appearance of the jet.  
Coincident with the flux maximum of 
knot B (see bottom left
panel of Figure 2), there is a significant minimum in optical
polarization of $\simeq 10\%$ (see Figure 8).  No change is seen,
however, in the optical magnetic field vector direction at this location. 
Just downstream of
this, the optical polarization increases to $\simeq 35\%$.   A second
decrease in optical polarization is seen to coincide with 
the upstream edge of the dust ring, along the jet's ridgeline.  
Also beginning at about 0.5'' from the core, the optical magnetic field vectors 
are seen to rotate by about 30 degrees to assume more of a north-south 
orientation, particularly along the northern edge of the jet.
In the optical, at the jet edges the level of polarization increases to upwards
of $\sim 50\%$, and the projected B-field tends to be less aligned with the jet
direction. 

\begin{figure}
\label{Xsections}
\centerline{\includegraphics*[width=3.5in]{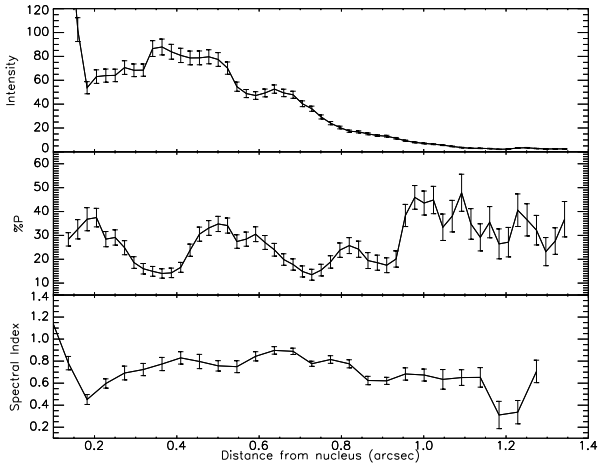}}
\caption{Cross sections along jet of ({\it top}) Intensity, ({\it middle}) \%P
and ({\it bottom}) optical spectral index.  Note the clear polarization minimum
at the jet intensity maximum between $0.3\sim0.4''$ from the core.}
\end{figure}

\begin{figure*}
\label{F330Wpol}
\centerline{\includegraphics*[width=6.5in]{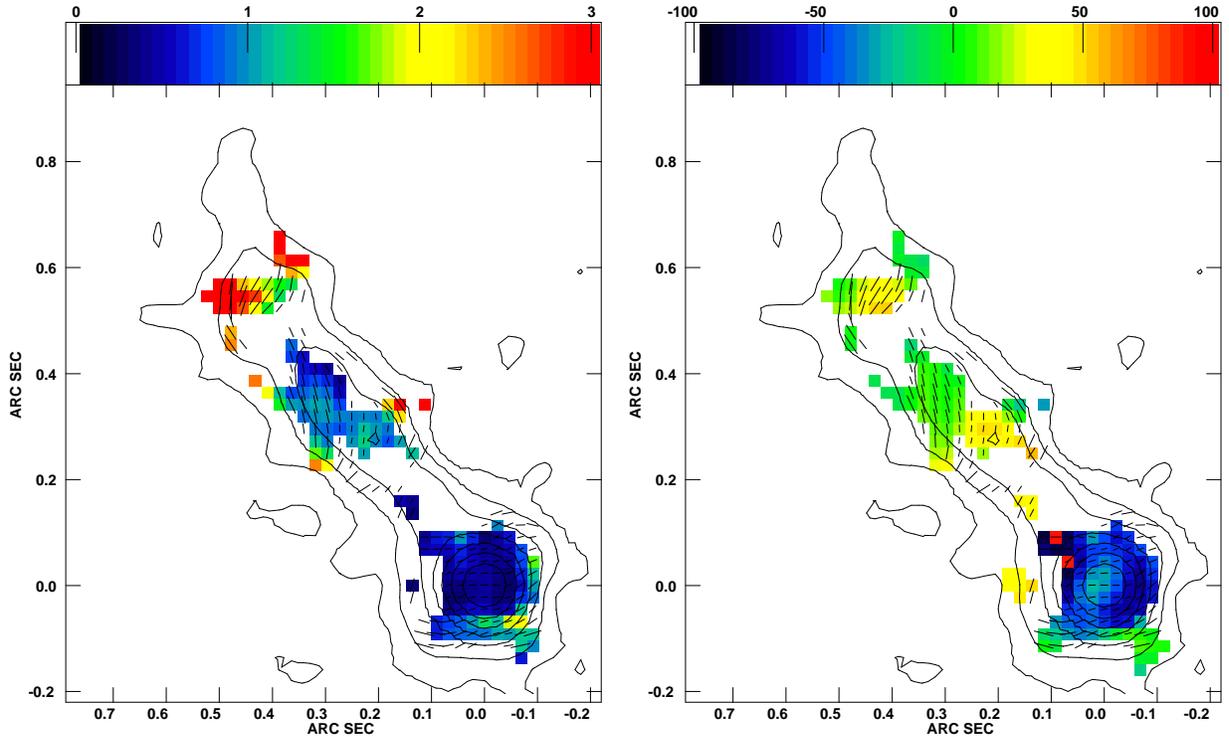}}
\caption{The high-resolution ACS F330W polarization data.  At left,  the color 
scale  shows the ratio of F330W polarization
to that seen in F555W at lower resolution, while at right, we show in color 
the difference in position angle.  In both panels we show contours in F330W flux
along with polarization vectors taken from those data.  Flux contours are 
plotted at 4.571$\times 10^{-2}~ {\rm counts/s} ~ \times$ ~ (1, 2, 3, 4, 6, 8, 12, 16,
24, 32, 48, 64, 128, 192, 256, 512, 1024).  The vectors represent B-field 
direction, and a vector 0.1 arcseconds long represents 100\% polarization.  
Note that the F330W data has a resolution more than a factor
2 bettern than the F555W data, thanks to its shorter wavelength and smaller pixel
size.  Thus despite its significantly lower S/N (a product of its much shorter
exposure time) it reveals the highest resolution detail in the polarization
structure of the optical/UV emitting part of the 3C 264 jet.  All of the differences
seen in this image as compared to Figure 4 can be attributed to the higher 
resolution.  See \S 3.2 for details. } 
\end{figure*}

The F330W ACS HRC polarization data (Figure 9), while much
lower signal to noise (thanks to the shorter exposure time), sheds light
on some regions of the inner jet of 3C 264, thanks to its higher resolution.  In
particular, we see that the knots in the inner jet do, in fact, have significant
polarization PA changes in the optical.  In particular, in knot A,  we see a low
polarization region towards the western side of the knot maximum,  continuing
through the ridgeline and then winding around to the eastern side  of the
knot B maximum.  Where we do see significant polarization in knot A, on the
eastern side, it has an orientation $\sim 90^\circ$ from the jet direction. 
Patterns of this type have been seen in other jets, such as in the knot C 
region of the jet of 3C 15 (Dulwich et al. 2007).   
Knot B's maximum is seen to have a
north-south magnetic field configuration, which persists into the B/C interknot
region.  Finally, in knot D, we see a zero-polarization region
along the ridgeline, with a higher-polarization region downstream where the
magnetic field vectors become perpendicular to the jet direction in the jet
center, and parallel to the jet direction along the eastern and western edges of
the component.  This  pattern  is similar to that seen in some features in the
M87 jet, in particular  its knot D (Perlman et al. 1999).  We will comment on
these issues in greater  detail, along with their physical implications,  in
Section 4.

Many similar features are seen in the K-band radio polarimetry data (Figure 3),
however
some differences also appear.  In Figure 10, we summarize the
differences and similarities in polarization properties between the
optical and radio polarization data.  We see broad similarities in several 
regions of the jet.  In particular, 
the radio data show a similar level of fractional polarization, when
averaged along the jet, as well as the same general trend in the orientation
of the magnetic field vectors.  We also see a similar increase
in the fractional polarization towards the jet edges, as well as a
change in the orientation in the magnetic field vectors towards a more
northern direction.  This latter trend at first blush appears to be
correlated with some differences in the $\Delta PA$ map shown in the
right panel of Figure 10; however, one must recall the
$180^\circ$ degeneracy that is inherent in polarization data, and
indeed closer inspection of both datasets reveal that this is the
nature of the red ``band'' apparent in that panel at this location.  This 
region is, however, seen to have a somewhat lower polarization in the 
optical than in the radio.

\begin{figure*}
\label{fig}
\centerline{\includegraphics*[width=6.5in]{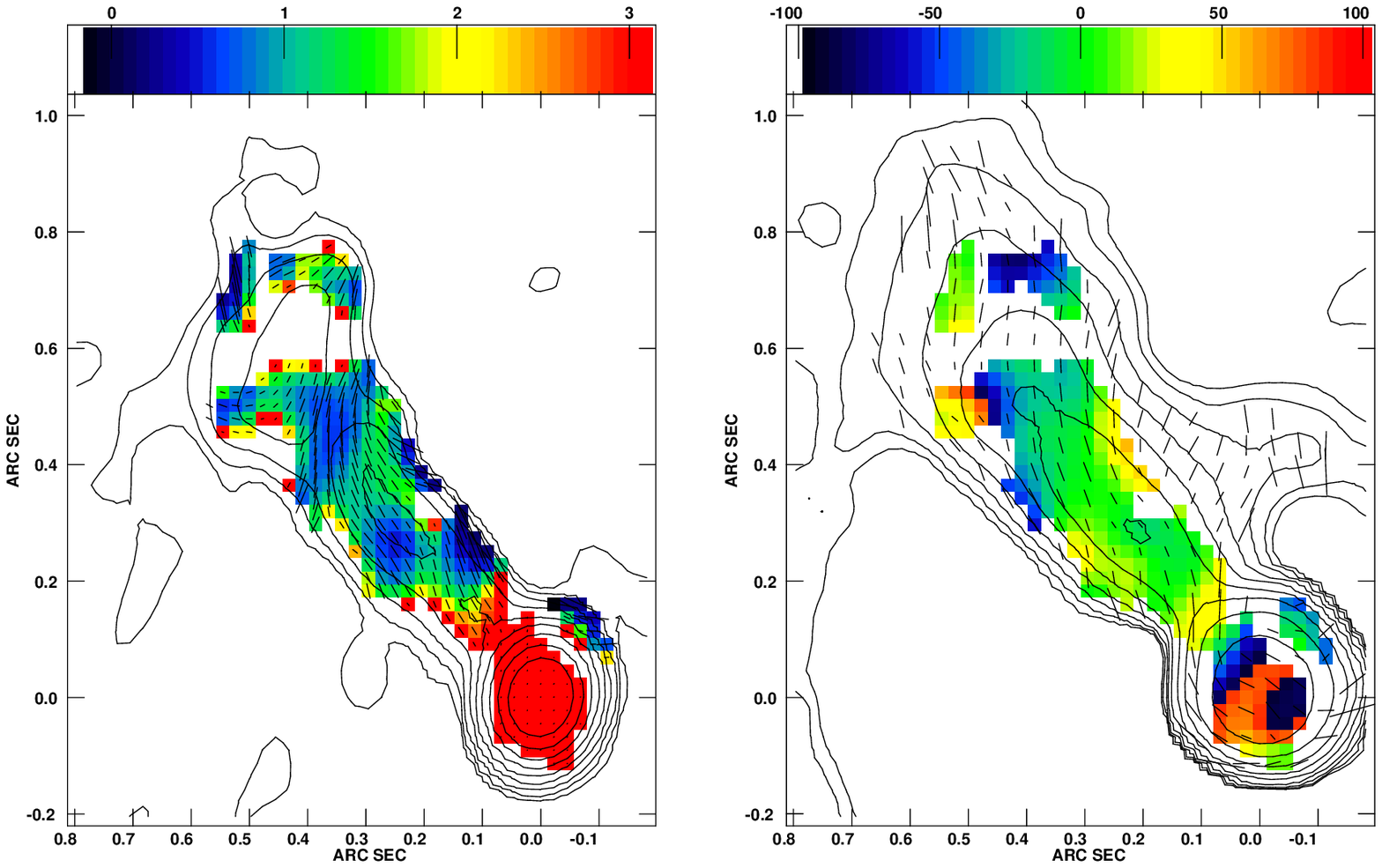}}
\caption{The comparison between Optical and radio polarization in the 3C 264
jet.  At left, we show the 22.5 GHz, A-array 
polarization map of Lara et al. (2004) (contours and polarization
vectors), overplotted onto a color image showing the ratio of optical to radio
polarization, while at right, we show the F555W polarization data (contours
and polarization vectors, but with vectors shown only every other pixel) 
overplotted onto a color image showing the difference
in polarization PA between the optical and radio data.  In the panel at left, 
contours are plotted at 4.522 $\times 10^{-3} ~ \times
(1, 2, 4, 8, 16, 32, 64, 128, 256, 512, 1024)$ Jy/beam, whereas in the panel at
right, contours are plotted at 2.529 $\times$ (1,2,4,8,16,32,64,128,256,512,1024)
counts/s.    As can be seen, there
are several regions where significant differences are seen between the degree
of fractional polarization seen in the optical and radio, but the differences 
in polarization position angle are small (note that the one region where there
appear to be some differences is very close to $180^\circ$ and hence 
degenerate.}
\end{figure*}

Upon closer inspection of Figure 10, however, we see several more significant
differences between the optical and radio polarimetry.  In particular, near the
maximum of knot B, where a minimum was seen in the F555W
polarization data, we see that the radio data do not show a similar
feature (see Figure 10, left panel).  Instead, we see in the radio map just 
upstream of that 
region (coinciding with knot A and upstream), 
a decrease in the radio polarization, to a value of only $\sim 5\%$, 
in a band that extends across 
the entire jet.  A closer look at the F330W data, however, does reveal a 
somewhat similar feature in this region, but it is still considerably 
narrower in the optical, where it is seen to stretch east-west across the
jet from west of knot A to east of knot B.
The knot A-B region exhibits similar
orientations of the polarization vectors in the optical and radio.
Even stronger differences
are apparent in a region $0.7-0.8''$ from the nucleus that extends across 
nearly the entire jet width.  In this region, which corresponds to the 
peak of knot D, no signficant polarization 
is detected in the K-band data, while in our F555W data the polarization is also
somewhat decreased, but higher, between 10-15\%.  This region, the downstream
end of which coincides
roughly with the location of the dust ring's upstream edge, also marks the
location of a subtle change in the orientation of the optical magnetic field
vectors to a north-south orientation. A look at the higher-resolution, but lower
signal to noise F330W data does reveal a low optical polarization region right
at the peak of knot  D, and downstream magnetic field vectors that seem to
surround the component, running
parallel to the local flux contours, including perpendicular to the jet 
direction at the ridgeline.    The low optical polarization region
is seen in the F555W data 
to continue out to the far edge of the dust ring.  This could be a result of
scattering due to the dust in the ring.  
These features will be discussed in more detail in \S 4.
%

On larger angular scales (e.g., beyond the dust ring), the optical surface 
brightness of the jet
decreases precipitously.  Nevertheless, we detect significant optical
polarization in this regions, with $P=30-40\%$ in a large region near the 
northern edge of the jet, extending between distances $\sim 0.9-1.5''$ from the
core.  In this region, the magnetic field vectors appear to be aligned
approximately with the jet, although the error bars and scatter on them are
large. The same region shows a similar radio polarization (see Figure 5 of Lara
et al. 2004), and polarization position angle (note that Lara et al. plot {\it
E} vectors rather than implied {\it B}, as we do). We do not have the signal to
noise to follow the polarization of the jet beyond this point, in the region
where the jet is seen to bend through 90$^\circ$, where the 5 GHz map of Lara et
al. shows an  east-west magnetic field vector and a somewhat smaller
polarization.      

\subsection{X-ray results}

Our deep Chandra observations (Figure 5) 
reveal X-ray emission 
from one and perhaps two regions of the jet, as well as the AGN itself. 
One of those two regions lies well beyond the optically bright part 
of the jet (at $>1.5''$ from the core), while the other is on scales less than
$1''$, and thus very close to the 
size of the Chandra HRMA PSF.  
Therefore it was necessary
to be very careful with our treatment of the {\it Chandra} observations.  This
entailed not only the use of SER and deconvolution techniques (described in \S 
2.3), but also exacting statistical tests.   

We first generated a simulated PSF with ChaRT (Carter et al. 2003) using 
the total spectrum extracted from the source (see below) 
as input for the ray-tracer, as well as the
source position found from centroiding. We characterized the PSF asymmetry in
the X-ray image
along the optical jet axis ($\sim 40^\circ$ East of North), by extracting
source counts from the simulated PSF in symmetric regions on both the jet and
counter-jet sides of the core.  
The asymmetry in the PSF favors the counter-jet direction
and was found to be approximately $30\%$.  We use the asymmetry estimate as
a systematic error term when calculating the net jet counts, which we added in
quadrature to the propagated statistical errors.
Next, we extracted background subtracted radial profiles spaced by $\sim 0.15''$
in the jet and counter jet directions from the sub-sampled SER event file.
Before extracting these profiles, we excluded all events outside of a $1''$
wide strip along the jet direction. 

By directly comparing the jet to counter jet profiles we were able to detect a
significant excess of counts on the jet side in the {\it Chandra} image, between
0.7$''$ and 1.8$''$ from the core.  This excess is shown in Figure 11.  The jet 
component appears to have an approximately flat (within the errors) surface
brightness profile  at these radii.  Thus while the jet is clearly resolved in
the X-rays, it is not possible with these data to comment in any more detail on
the structure of the X-ray emission between $0.7''-1.8''$ from the nucleus of
3C 264.  Note that while jet emission is likely detected as close as $0.5''$
from the  AGN, it is much harder to quantify the jet emission inside of $0.7''$
due to the  presence of the much brighter nucleus.  This latter region,
unfortunately,  coincides with the innermost of the two regions that the eye
picks out in Figure 6. 

\begin{figure} \label{XrayJet}
\centerline{\includegraphics*[width=2.75in,angle=270]{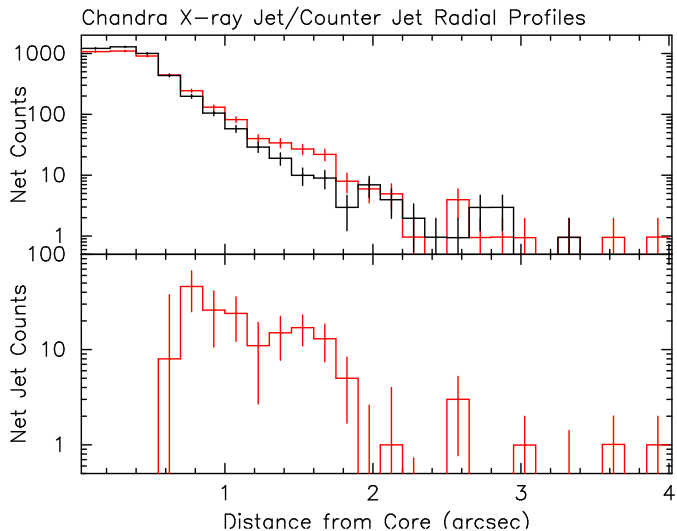}}
\caption{{\it Top}: Chandra X-ray count profiles along jet (red) and counter jet
(black). {\it Bottom}: Counter-jet profile subtracted from jet profile. Note the
significant excess between $\sim0.8''$ and $\sim 1.8''$.} \end{figure}


The 0.3--10 keV X-ray spectrum of the core, containing 10584 net counts and
binned to a minimum of 30 counts per channel, is found to be well fitted by an
absorbed power law with $\alpha = 1.24 \pm 0.05$  (1-$\sigma$ error bars
for two interesting parameters;
$\chi^2_{\rm min}$ = 138 for 152 degrees of freedom).  As shown in
Figure~\ref{fig:corespec}, the absorption is consistent with the Galactic value
of $N_{\rm H} = 2.45 \times 10^{20}$ cm$^{-2}$ (from the COLDEN program provided
by the {\it Chandra\/} X-ray Center, using data from \cite{dlock90}).  The
intrinsic luminosity over the energy band 0.5 -- 8 keV is $1.6 \times 10^{42}$
erg s$^{-1}$.  Results are in agreement with Donato et al.~(2004: 
{\it XMM-Newton})
and Evans et al.~(2005: {\it Chandra\/}) when the same energy bands are
compared.

\begin{figure}
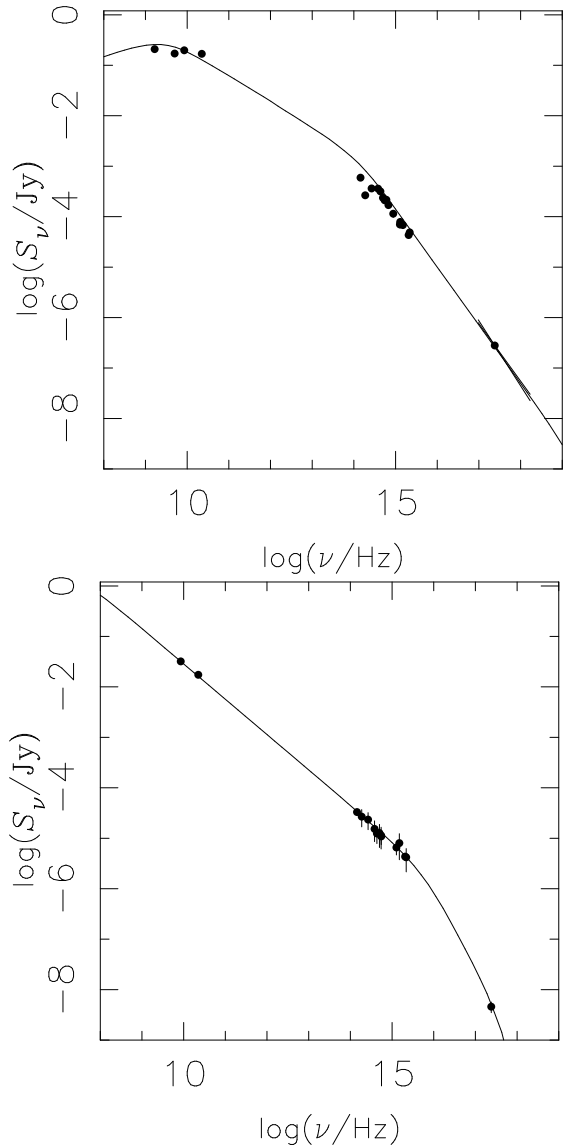

\plotone{3c264-break.ps}
\plotone{3c264-jet.ps}
\caption{{\it Top:} Spectrum of the nucleus using data from this paper
and radio points at 1.66 and 5~GHz from Lara et al. (2004).
A simple one-zone synchrotron model (solid line) can describe the overall
characteristics of the data as arising from an unresolved small-scale
jet.  {\it Bottom:} Spectrum of the outer jet also fitted to a
synchrotron model.  See text for details.
\label{fig:radiotoxray}
}
\end{figure}

While the core is so bright that the choice of background region on
the chip makes little difference to its fitted X-ray spectrum, much of
this `background' is from the cluster A1367, within which
3C~264 resides.  There is also evidence for thermal emission on a
smaller spatial scale associated with the gaseous atmosphere of
NGC~3862, the host galaxy of 3C~264. This was first noted by Donato et
al.~(2004) using {\it XMM-Newton} data, but a measurement of the gas
properties was limited by the larger point-spread function of XMM-Newton
and the fact that the source was far from
the optical axis of the telescope.   The first {\it Chandra}-based 
study of A1367
was done by Sun et al. (2005); however, that work did not discuss 3C 264
in depth as it was carried out with an earlier dataset centered on the cluster
center, which is $8'$ distant from 3C 264.    
Sun et al. ~(2007), using the same {\it
Chandra\/} dataset observations as herein, 
measured an excess of soft X-ray counts compared with
the PSF in an annulus of radii $1.5''$ and
$6''$ (0.7--2.6 kpc) centered on the nucleus of 3C~264.  They found
$kT = 0.65^{+0.29}_{-0.09}$ keV for this emission.  We find some
evidence for the presence of gas even in our core spectrum, with
$\chi^2$ decreasing by 11 if such a component is included.  The
temperature is cooler than found by Sun et al.~(2007), at $kT =
0.30^{+0.11}_{-0.06}$ keV, indicative of complex thermal structure in
the galaxy.  The inclusion of the thermal component does not
significantly affect the spectral index of the dominant power-law emission, 
however (with a thermal component we obtain $\alpha=1.18 \pm
0.04$). Its normalization is reduced by about 5\%, but since this is roughly
the same as the percentage of missing flux due to the wings of the PSF, the
normalization in Fig.~\ref{fig:corespec} is a good
representation of the 1~keV flux density of the nucleus ($0.28 \pm
0.01$ $\mu$Jy).

Following the method described in \S 2.3, the net count rate in the
resolved jet is $(4.5 \pm 1.1) \times 10^{-3}$ count s$^{-1}$.
Adopting the spectral index of the core gives a 1~keV conversion
factor of 1028 nJy count$^{-1}$ s (0.5--8 keV), similar to the canonical value of
1000 nJy count$^{-1}$ s of Marshall et al.~(2005), and thus a 1 keV flux
density for the resolved jet of $4.6 \pm 1.1$ nJy.
This flux only applies to the jet outside of $0.8''$ from the
core, for the reasons discussed above. We should also note that if the jet 
does in fact extend into the core, as
the optical and radio jets do, then the core spectrum discussed above will be
contaminated by jet components. 
Assuming that the X-ray flux beyond $1''$ 
is indeed due to the jet, we find an optical to
X-ray spectral index for the outer jet of $\alpha_{OX} = 1.48 \pm 0.06$.
Combining this with our fitted radio to optical spectral index for the outer
jet of $\alpha_{RO} = 0.723 \pm 0.003$, we can estimate the minimum spectral
break frequency and the minimum break in $\alpha$. Doing so, we find (again,
for the jet outside of $0.7''$) $\nu_{break} \gsim 2 \times 10^{15} {\rm ~Hz}$,
and $\Delta\alpha \gsim 0.85$. 

\begin{figure}
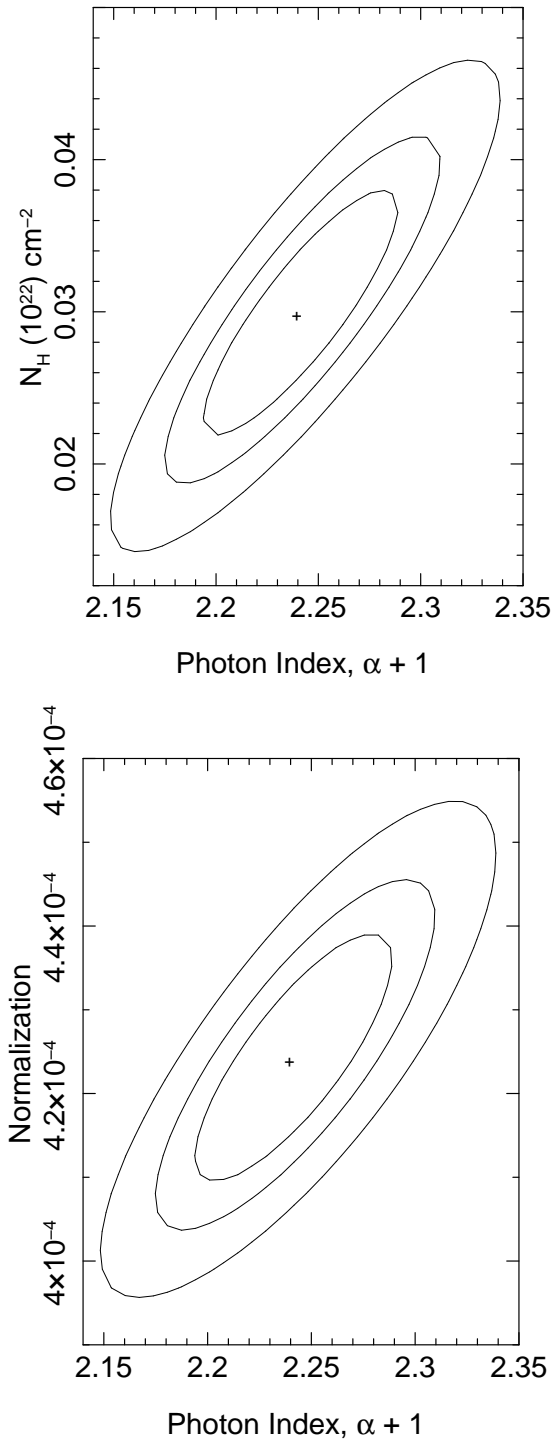


\plotone{core-pl-nhindex.ps}
\plotone{core-pl-normindex.ps}
\caption{Parameter uncertainty contours for the X-ray spectrum of the 
core fitted
to an absorbed power law. Top:
spectral index against $N_{\rm H}$.  Bottom:
spectral index again normalization in units of cm$^{-2}$ s$^{-1}$
keV$^{-1}$.  $\chi^2$ contours are 
1$\sigma$, 90\%, 99\% for two interesting parameters. See \S 3.3 for details.} 
\label{fig:corespec}

\end{figure}

\section{Discussion}

In the preceding sections, we have discussed in detail optical and radio 
polarization images of the jet of 3C 264, as well as its optical and broadband
spectrum and the results of a deep {\it Chandra} X-ray observation.  As already 
discussed, this jet displays a smooth morphology in both optical and radio.
The differences we see in polarization morphology, while less drastic than 
those seen in e.g., M87 (Perlman et al. 1999), are nevertheless significant.
We also find significant variations in the optical and radio-optical spectrum 
of the jet.  X-ray emission is also seen from the 
jet on scales between $0.7-1.8''$ from the nucleus.
In this section, we discuss the inter-relationship of these properties and 
their relevance to the overall physics of the 3C 264 jet.  We will also 
discuss other jets where radio and optical polarimetry has recently been done,
in particular M87 (Perlman et al. 1999, 2001, Perlman \& Wilson 2005), 
3C 15 (Dulwich et al. 2007), and 3C 346 (Worrall \& Birkinshaw 2005, 
Dulwich et al. 
2009), and their relationship to this source.

\subsection{The Physics of the 3C 264 System}

3C 264 is a highly interesting radio galaxy that contains many elements that 
are important to understanding both the AGN phenomenon as well as jet physics.
The data discussed herein allow us to place significant new constraints
on the physics of the nucleus and jet, and also to comment on the 
nature of the host galaxy environment.  The latter is quite interesting in
and of itself, as the host galaxy to 3C 264, namely NGC 3862, lies within -- 
but not at the center of -- a rich cluster that has complex substructure.
Cortese et al. (2004) have found that Abell 1367, the surrounding cluster, 
lies at the intersection of two filaments within the Great Wall, and is 
likely in the midst of its formation process.  The cluster has a large number
of subclumps and grouping, as revealed by multi-object spectroscopy and 
X-ray studies (Cortese et al. 2004; Sun et al. 2005, 2007).  NGC 3862 itself
lies at the center of a compact group within the cluster that consists of
18 galaxies (Lipovka et al. 2005).  Our {\it Chandra X-ray} data reveal that
the complicated temperature structure of the A1367 system continues down to 
scales of kpc, with evidence of a cooling-flow  like structure within the 
subgroup.

NGC 3862 was dubbed ``optically dull'' by Elvis et al. (1981), referring to the
relative strength of nuclear X-ray to optical emission.  Further investigation
of galaxies with similar characteristics (sometimes called X-ray bright
optically normal galaxies, or XBONGs) have normally found that they can be
fitted into the scheme of conventional accretion models if there is a moderate
amount of nuclear gas ($N_{\rm H} \sim 10^{21}$ cm$^{-2}$) and dust (E(B-V)
$\sim 0.5-0.8$) covering large solid angles (e.g., Civano et al. 2007). 
While our
HST images do show a dusty disk a few hundred parsecs from the AGN, this disk
does not appear to continue all the way to the center, and indeed, we find no
evidence of reddening or absorption of the nucleus, based both on our HST and 
{\it Chandra} data.  A similar conclusion was reached by Hutchings et al. (1998)
based on analysis of the nuclear spectrum. Thus this distinct lack of 
nuclear gas makes 3C 264 rather different from most XBONGs.  However,   
in 3C\,264 there is a bright radio nucleus, and
the radio and X-ray emissions could be enhanced relative to the
optical continuum if the continuum nuclear emission\footnote{We note
the presence of narrow forbidden lines in the optical spectrum
(Hutchings et al. 1998) } is primarily of a non-thermal origin.        

Our data make it clear that the primary radiation mechanism of the 3C 264 
jet and nucleus in the radio through ultraviolet is synchrotron emission.   
A broken-power-law synchrotron model (left panel of
Fig.~\ref{fig:radiotoxray}) describes the overall characteristics of
the radio-to-optical spectrum of the nucleus rather well, especially
given the non-contemporaneous nature of the data.  We have assumed a
spherical blob of radius 2~pc, in which an electron spectrum of
minimum Lorentz factor $\gamma_{\rm min} = 300$ produces radiation of
spectral index $\alpha = 0.5$ in a minimum-energy magnetic field of
$\sim 9$~mG, breaking by $\Delta \alpha = 0.7$ at $\gamma_{\rm break}
= 7 \times 10^4$ (see e.g., Worrall \& Birkinshaw 2006 for relevant formulae).
The dominant X-ray inverse Compton mechanism is synchrotron
self-Compton, but this gives a negligible contribution to the observed
X-rays.  It is interesting that the synchrotron lifetime of electrons
producing emission close to the break frequency is $\sim 5$ yr, of the
same order of magnitude as the light-travel time across the source.
However, $\Delta \alpha$ is greater than the value of 0.5 expected for
a simple energy-loss model, and is in the range $0.6-0.9$ commonly
found for the resolved jets of low-power radio galaxies, and not yet
readily explained by acceleration models (see Worrall 2009 for a review).
We return to this latter issue in section 4.2.  
While a synchrotron model can describe the overall
features of the nuclear continuum of 3C\,264, we note that the more
extreme case of a radio and X-ray bright (and unabsorbed) radio galaxy
with an optically dull continuum, PKS~J2310-437 (Tananbaum et al. 1997,
Worrall et al. 1999), has proved to be more problematic to fit to a simple
synchrotron model (Bliss et al. 2009).

Also supporting a synchrotron nature for the nuclear emissions in the 
optical are the polarized emissions found by our HST observations.  A
similar conclusion was
reached by Capetti et al. (2007), who note an optical polarization $\sim 8\%$ 
for the core and a position angle for the electric field vectors $\sim
5^\circ$.  This fractional polarization is too high to be reached with
scattering (see Capetti et al. for discussion) and is most consistent with
synchrotron radiation.  It  should be noted, however, that the magnetic field
orientation seen by Capetti et al. (2007) is somewhat different (by about 25
degrees)  from that seen either in the K-band, A-array data we (or Lara et al.
2004) present, and is also different from that seen in most of  the VLBI jet
(Kharb et al. 2009).  Unfortunately we cannot use our {\it WFPC2}  polarimetry
data to comment on this latter issue, as the core is saturated.  

The optical emissions from the jet have been known to be due to synchrotron
radiation for more than a decade (e.g., Baum et al. 1997).  Our data 
allow us to extend this conclusion considerably.  Not only do they allow us
to place much better constraints on the shape of the jet's optical SED, but 
they also show that synchrotron radiation is the dominant jet emission 
mechanism up to X-ray energies.  
While the X-ray spectrum of the resolved jet is not measured, the
emission is undoubtedly of synchrotron origin.  X-rays are measured
only from the outer jet where contamination from core emission is
least, but they arise from a relatively large region.  We have
modelled this section of jet with a cylinder of radius 20~pc
\citep{Lara04} and length 550~pc, and used $\gamma_{\rm min} = 300$.
Again inverse Compton scattering, including on the cosmic microwave
background radiation, gives a negligible contribution to the observed
X-ray emission.  Again a synchrotron model of a broken power law fits
well the data (right panel of Fig. 13).  To find a
minimum value of $\Delta \alpha$ we allow the break to occur at as low
a frequency as possible, but as for the core the value of $\Delta
\alpha$ (0.82) is greater than 0.5, and in the range $0.6-0.9$
commonly found for the resolved jets of low-power radio galaxies for which
the X-ray spectrum is measured.  The jet spectral index below the
break is $\alpha=0.7$, the minimum-energy magnetic field is $\sim 330
\mu$G, and the Lorentz factor of electrons emitting radiation close to
the break is $\gamma_{\rm break} = 1.6 \times 10^6$.  The synchrotron
lifetime of electrons producing radiation at the break frequency
($\sim 140$ yr) is again of the same order of magnitude as the
light-travel time across the source.

As noted above, the jet of 3C 264 shows rather subtle magnetic field direction
changes, both in the optical/UV and radio.  The first region where changes are
seen to occur are the knot A/B region, which displays a low polarization region
in the interknot region between them, that extends northwest  along the
northern  side of knot A as well as northeast along the eastern side of knot
B.   Twists in the 30-90$^\circ$ range are seen to the side of this feature.
The low polarization at the knot maximum is likely to be the result of field
superposition, and the polarization angle changes may be due to the occurrence
of perpendicular and oblique shocks at the location of the knot maxima.
A similar pattern of changing position angles in knot C of the 3C 15 jet has,
alternatively, been modelled as the superposition of a helical strand of
magnetic field on a more ordered field structure. 
We do not have the angular resolution in our X-ray data
to see anything that might be related to the latter feature (as we do in 3C 15), 
however, a comparison of the two sources and their
associated maps reveals some similarity.   A second region is seen with
significant rotations in the orientation of the magnetic field, namely the knot
D region.  This region shows low polarization in the jet  interior and
particularly at the knot maximum, with magnetic field vectors that  follow the
jet's flux contours downstream.  Such a configuration is very similar to  some
of the knots in the M87 jet, particularly knots D and F.  Perlman et al. 
(1999) modeled those features as shocks that occur within the jet interior, in
a  stratified jet where the optical/UV  emitting particles are also 
concentrated in
the  jet interior, while lower energy, radio emitting particles are
more evenly distributed through the jet.  Such a model can be
applied  for the most part to these knots as well, with the possible exception
that here, we do not have sufficient evidence in our spectral index maps to say
that the knots are much more optically bright than the edges, although the
$\alpha_{ro}$ map does show that they do have a slightly  flatter radio-optical
spectrum.   Thus, as in the M87 system,  the knots in the 3C 264 jet 
regions are likely to be  sites of enhanced {\it in situ} particle acceleration.


The nature of the dust ring in 3C 264 and its relationship to the jet is an 
interesting topic.  We see some apparent steepening in the jet's optical  
spectrum (from $\alpha_O \approx 0.75$ to $\alpha_O \approx 0.9$) 
near this location, as well as significant changes in polarization in
both the optical and radio. These changes include minima in the radio
polarization and changes by 20-40$^\circ$ 
in the direction of the magnetic field vectors
in the optical.  The former does not  necessarily indicate an interaction
between the jet and the dust ring, but rather is more consistent with absorption
of jet radiation   by the dust within the ring.  Moreover, a closer look 
at Figure 4 reveals that the spectral and polarization
changes begin about $0.1''$ closer to the nucleus than the inner edge of the
ring.   In addition, 
the changes in the direction of the optical polarization vectors are
much stronger near the  northern edge and continue in the region beyond the
ring.  This indicates that the spectral and polarization changes are primarily 
intrinsic to the
jet rather than caused by any screening by or interaction with the ring.  Thus,
we believe that
the disk and jet most likely occupy physically distinct regions of 
the galaxy.  A similar conclusion was reached by Martel et al.
(2000) on the basis of the continuous distribution of line emission at the 
location of the disk, with the galaxy subtracted long-slit spectrum of 
Hutchings et al. (1998) showing no evidence of line emission from the jet 
itself.

Beyond $1''$, the optical polarization is much harder to measure, due to the
much lower optical flux.  Interestingly, however, where it is measurable, 
we see an increased  optical
polarization in this region, as high as 40\%, with an orientation 
slightly different than that seen in the rest of the inner jet, perhaps reflecting
the turn towards a more northerly orientation that is seen in the radio flux contours
at  slightly greater distances (Figures 2, 3, 8
and Lara et  al. 2004). 

The overall morphology of the radio through X-ray spectrum of the core+jet
(Figure 7, {\it bottom right panel}), combined with the power-law slope of the
optical-to-X-ray spectrum, is most consistent with synchrotron radiation.  
Interestingly, the  spectral index observed for the core + inner jet in the 
X-rays is reasonably similar (within $2 \sigma$) to  that observed in our data
in the ultraviolet for the entire jet, as well as that of 
the inner jet (the outer
jet is too faint to see in the UV data).  This implies little curvature of the
spectrum between the UV and  X-rays, and also implies that the $\nu F_\nu$
peak of the jet's spectrum is located at a few  $\times 10^{14}$ 
Hz, over a broad
range of jet components.  The spectral break required for the
inner 3C 264 jet  is $\Delta \alpha = 0.6$; slightly but not significantly
smaller than seen in 
M87 ($\Delta \alpha =
0.7$, Perlman \& Wilson 2005).   Also in accord with the  synchrotron radiation model, we see that the
width of the jet is very similar in the radio, optical and UV, for as long as
the data allow us to track it (Figure 14). Interestingly, the jet FWHM increases
smoothly with distance from the nucleus in  all three bands, with the width $W
\propto r^{1.1}$ approximately.  

\begin{figure}
\label{width}
\centerline{\includegraphics*[width=3.5in]{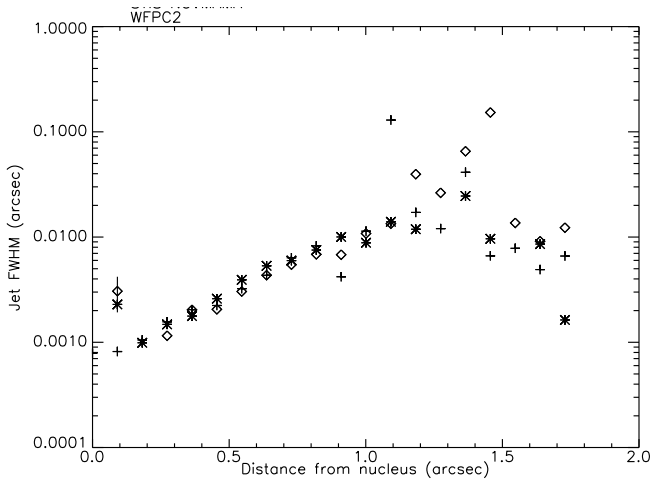}}
\caption{The width of the 3C 264 jet in the UV, optical and radio.  Three datasets
were used: The VLA K-band, A array data (crosses), Our Stokes I, F555W data
(diamonds), and the NUVMAMA image (asterisks).  Note the constant rate of increase
in jet width, as observed in all three bands out to about $1.2''$ from the core.}
\end{figure}

\subsection{Relationship to Other Jets}

It is useful to discuss these results in the broader context of the class of 
optical and X-ray jets.  Ten such jets have now been observed  polarimetrically
with HST:  M 87 (Capetti et al. 1997, Perlman et al. 1999), 3C 273 (Thomson et
al. 1993), 3C 293 (Floyd et al. 2006), 3C 15, 3C 66B, 3C 78, 3C 264, 3C 346, 3C
371 (Perlman et al. 2006; see also Dulwich et al. 2007, 2009)  and most recently
PKS 1136-135 (Cara et al., in prep.).  Of these the vast majority (all but 3C
273 and PKS 1136-135) are relatively low-power sources, albeit with a variety of
optical and radio morphologies. These objects are all at a wide range of
distances, ranging from just 16 Mpc in the case of M 87 to $z=0.534$ in the case
of PKS 1136-135.  Of these ten, just three (M 87, Perlman et al. 1999; 3C 15 and
3C 346; Dulwich et al. 2007, 2009)  have previously had deep analyses of their 
radio and optical total flux and polarization morphology, as well as broadband
spectra and X-ray emission published (importantly, note that the existing  HST
polarimetry of 3C 273 is not deep enough for these purposes).   Compared to the
other objects, 3C 264's  optical and X-ray jet is the most compact:  at $2''$
its angular length is less than any other optical/X-ray  jet for which
polarimetry has been done, except for 3C 78, which is very similar in length,
although its physical length of about 850 pc (projected) is not that much
shorter than the 1.5 kpc projected extent of the M87 jet. This limits our ability
to extend the detail we can obtain in the  radio and optical imagery to the
X-rays, as the jet of 3C 264 is only barely  resolved with {\it Chandra}.  Thus
with these observations we cannot comment in detail on the correlation between
changes in optical polarization and the  location of X-ray components,
previously noted by Perlman \& Wilson (2005),  Perlman et al. (2006) and Dulwich
et al. (2007, 2009).  

The dominant polarization configuration observed in the jet of 3C 264, in both
the radio and optical, is one of a high polarization (40-50\%) along the jet
edges, and somewhat lower polarization in the jet center, with magnetic field
vectors oriented primarily along the direction of the jet.  Such  a
configuration is seen in most of the knot A-B complex of the M 87 jet, albeit
with  considerably more detail because of the greater proximity of that
source.    What is missing in 3C 264, however (as compared to M87), are the
presence of  strong "shock-like" components, which are seen not only in the A-B
complex of the M 87 jet, but also commonly in the other two jets as well as
other regions of the M 87 jet.  By contrast, the jet of 3C 264 has a very
smooth polarization morphology, unlike any of the
other three jets analyzed deeply so far and similar only to 3C 78 in detail.  
Perhaps related to this, we note that 3C 264 and 3C 78 represent the lowest
power jets of those that have been observed polarimetrically with HST. It
is therefore possible that the overall character of the polarization morphology
is partly a function of the  jet power.  We will test this in our future studies
of the remaining jets in the Perlman et al. (2006) sample (3C 78, 3C 371 and 3C
66B,  for which future papers are all under preparation). 

We turn now to an effect observed in all four jets: the spectral energy
distribution (SED) peaks at $\sim $ optical-UV energies, and is followed by a
power-law extension, with a break steeper than the canonical $\Delta\alpha=1/2$.
In M87,  the spectrum appears to continue to steepen at X-ray energies; however
this is not the case in 3C 264.    Although a possible formal  explanation for
this steeper break is that electrons do not diffuse in pitch angle as they
radiate (Kardashev 1962, Pacholczyk 1970), a more plausible explanation is that
the magnetic field decreases away from the sites of particle acceleration. This
naturally  results in spectral breaks greater than $1/2$, assuming that
electrons, after they are accelerated propagate toward  regions with lower
magnetic field (Wilson 1977). Recent work by Reynolds (2009) has expanded upon
this aspect, associating steeper breaks with inhomogeneities and/or differential
losses within the source. 
Another possibility that can work in parallel with the magnetic field gradients
is that of velocity gradients in the flow.  This idea was first posited by Laing
and Bridle (Laing 1996, Bridle 1996) to deal with observed details of radio 
polarization maps of jets, which suggested a spine-sheath configuration for 
the jet, with a faster flow in the central spine, similar to a firehose.  
Later work invoked lateral
velocity gradients to model the sub-pc scale jets of TeV emitting
BL Lacertae objects (Georganopoulos \& Kazanas 2003), and can be also applied to
transverse velocity gradients that may characterize the kpc scale jets of M87
and 3C 264.  A velocity gradient across the jet would effectively reduce and/or 
inhibit a field component in the direction perpendicular to the velocity
gradient, and thus would allow for a different polarization morphology in 
the jet center, as observed here.  In this scenario,  the high energy emission 
would come primarily from the fast moving spine, while the low energy emission 
would be dominated by emission from the slower sheath (Ghisellini et al. 2005).  
For small angles $\theta<1/\Gamma_{sp}$, where $\Gamma_{sp}$ is
the bulk Lorentz factor of the spine, the observed cooling break is the
canonical $1/2$ break predicted by simple theory. At larger angles, however, the
high energy emission of the spine is beamed out of our line of sight more than
the sheath emission, and this results in spectral breaks steeper that $1/2$ (see
Figure 2 of Georganopoulos \& Kazanas 2003).   It is important to point out that
these two possible explanations are not mutually exclusive;  indeed, this can be
seen by the breadth of work that has developed on the SED of the M87 jet, where
the spectrum continues to steepen through the X-rays.  This caused Perlman \&
Wilson (2005) to invoke particle acceleration in an energy-dependent part of the
cross-section of the jet to explain this, while other authors (Fleishman 2006,
Liu \& Shen 2007, Sahayanathan 2008) have  expanded on this by invoking 
two-zone, and in one case a diffusive, shock models.  Work on other jet sources
has expanded upon this by invoking velocity shear and deceleration (NGC 315: 
Worrall et al. 2007, 3C 31:  Laing et al. 2008;  see also Wang et al. 2009).
In these sources, both toroidal and longitudinal fields are needed to account
for details of the polarization morphology at the jet edges.  These last two
sources are rather similar in flux and polarization morphology to 3C 264, 
albeit longer and without optical emission from their jets.  This is
particularly true when the details of the F330W polarization maps (Figure 9)
are taken into account.
Note, however, that while the work to date suggest that particle acceleration  
is enhanced in shock-like knots (e.g., Perlman \& Wilson 2005), a continuous 
mode of particle acceleration also appears required for most X-ray jets (see
e.g., Perlman \& Wilson 2005; Jester et al. 2001, 2005, 2006; 
Hardcastle et al. 2003, 2006, 2007).

Another issue is orientation.  For M 87, this is
reasonably well constrained now, as the superluminal motions (Biretta et al.
1995, 1999, Cheung et al. 2007) seen in its jet restrict the angle to our line
of sight to values $\leq 18^\circ$ (although n.b., this is in conflict  with
estimates derived from the observed geometry of the disk and knot A, e.g.,
Ford et al. 1994).  A similarly small viewing angle is seen in the 3C 346 jet,
which Dulwich et al. (2009) has constrained to be at an 
orientation of 14$^{+8}_{-7}$ degrees, before bending 22 degrees at an oblique
shock several arcseconds down the jet.  This value is consistent with earlier 
VLBI analysis as well (Cotton et al. 1995).  By comparison,      for 3C 264, 
modeling  suggests an orientation angle as large as $\theta\sim 50^\circ$ (Baum
et al 1997, Lara et al. 1997), based on the jet to counterjet ratio as well as
VLBI and ROSAT data.  No firm constraints exist on the orientation of the 3C 15
jet, the only other one for which a detailed comparison of radio  and optical
polarimetry has been published to date (Dulwich et al. 2007). As was argued by
Perlman \& Wilson (2005) in the case of M87 most of the synchrotron optical and
X-ray emission comes from the spine of the jet. If we assume that the high
energy emitting spine is also faster than the mostly radio emitting sheath, a
result supported by numerical simulations of relativistic jets (Aloy et al
2003), then for sources at greater angles to the line of sight, as
possibly is 3C 264, the high energy emission from the fast spine will be more
de-amplified  than the slower sheath emission. In this scenario, most of the
radio emission comes from the slower sheath, independent of orientation, and
this may explain the similar
radio polarization of 3C 264 and M 87 as shear due to transverse
velocity gradients in the plasma is likely to cause the magnetic field
to align with the jet.  The difference between highly aligned (e.g.,
M87) and less aligned (e.g., 3C 264) sources appears at higher energies.  In 
the case of highly aligned jets, one expects that the optical and X-ray emission 
we observe is dominated by the fast spine, where a significant part of the high
energy particle acceleration takes place, most probably in shocks manifested 
through magnetic field orientations perpendicular to the jet axis -- as indeed is
seen particularly in the polarization data for the inner knots, e.g., 
Perlman et al. (1999), Perlman \& Wilson (2005).
However, in 3C 264, which is seen at a considerably larger angle, any high-energy
emission coming from the fast spine will be de-amplified and thus the X-ray 
emission we see is most likely originating in the slower sheath.  This may 
explain the fact the optical polarization of 3C 264 is in general parallel
to the jet axis, following the radio jet polarization pattern.  It also suggests
that 3C 264 requires a mechanism for distributed particle acceleration, as
suggested also for the much more powerful jet of 3C 273 by Jester et al. (2007).  
This scenario needs testing with optical and radio polarimetry
of additional jets.  


\subsection {Summary}

We have presented new VLA, {\it Chandra} and {\it HST} imaging and polarimetric
data on the jet of 
3C 264 that allow us to make a comprehensive, multiband study of the energetics
and structure of this jet, and the relationship of these factors to the
high-energy emission.  We present polarization maps in the radio and optical, 
as well as total flux and spectral index maps in both bands.  We also present 
the first X-ray image of the jet and discuss its relationship to the structure
seen in other bands.  The dominant emission mechanism of the jet in the X-rays 
is found to be synchrotron emission.  
We find significant curvature of the jet spectrum both between the radio and
optical and also within the optical-UV band.
We present a model of the jet emission
that includes a smooth extrapolation of the optical-UV spectrum into the X-ray
with a  break at particle Lorentz factors $\gamma = 1.6 \times 10^6$ 
that is steeper than the canonical 0.5.   We discuss mechanisms for
producing the observed high-energy emission and the anomalously large spectral 
breaks observed in 3C 264 and M87, and we suggest that it can be due to
differential losses, velocity gradients
and/or magnetic field gradients coupled with more effective particle acceleration
taking place in the faster and/or higher magnetic field parts of the flow.  It is,
however, important to note that
a full test of these mechanisms would require higher-angular resolution data 
than currently available, as {\it Chandra} barely resolves the jet emission 
in the X-rays.   In the polarization data,  we find a 
smooth yet complex magnetic field structure in the jet which, while without 
exhibiting sharp, 'knotty' structures (as seen in some other high-resolution
polarization maps, e.g., M87), displays a structure that is correlated with 
changes in the optical spectrum.    While such a configuration is seen also in other jets, such
as M87, here the orientation is rather different, with
3C 264 being at a rather larger angle to our line of sight.  Thus in 3C 264 the
spine-sheath configuration manifests itself in the form of more subtle differences
between optical and radio polarization.   These include 
significantly higher polarization at the jet edges, as
well as twists in the magnetic field orientation near the positions of two knots
and in the interknot regions, seen particularly in the optical
polarization data in the knot A/B region and near knot
D.  Such a configuration is also consistent with the observed spectral energy
distribution of 3C 264, but requires a distributed particle acceleration mechanism.

\acknowledgments

Research on jets at FIT and UMBC is funded by NASA LTSA grants NNX07AM17G,
NNG05GD63ZG and NAG5-9997, as well as
NASA ATFP grant NNX08AG77G.  Other support for this work came from HST grants 
GO-9847.01, GO-9142.01 and {\it Chandra} grant SAO-05701071.  We thank an
anonymous referee for comments that significantly improved this work.


%


\end{document}